\documentclass[iop]{emulateapj}
\usepackage{amssymb}
\usepackage{graphicx}
\usepackage{grffile}
\usepackage{epstopdf}
\usepackage{url}

\newcommand{\msun}{\ensuremath{\rm M_\odot}}

\newcommand{\msunyr}{\ensuremath{\rm M_{\odot}\,{\rm yr}^{-1}}}
\newcommand{\secpoint}{\mbox{$''\mskip-7.6mu.\,$}}
\newcommand{\minpoint}{\mbox{$'\mskip-5mu.\,$}}
\newcommand{\Ha}{\ensuremath{\rm H\alpha}}

\newcommand{\lya}{\ensuremath{\rm Ly\alpha}}
\newcommand{\wlya}{$W_{\rm Ly\alpha}$}
\newcommand{\dvlya}{$\Delta v_{\rm Ly\alpha}$}
\newcommand{\Alya}{$A_{\rm Ly\alpha}$}

\newcommand{\kms}{km\,s\ensuremath{^{-1}}}
\newcommand{\ztwo}{\ensuremath{z\sim2}}
\newcommand{\zthree}{\ensuremath{z\sim3}}

\newcommand{\OIII}{[\ion{O}{3}]}

\newcommand{\HII}{\ion{H}{2}}

\begin{document}

\title{The \lya\ Properties of Faint Galaxies at $z\sim2$--3 with\\ Systemic Redshifts and Velocity Dispersions from Keck-MOSFIRE\altaffilmark{*}}
\author{Dawn K. Erb\altaffilmark{1}, Charles C. Steidel\altaffilmark{2},  Ryan F. Trainor\altaffilmark{2},  Milan Bogosavljevi{\'c}\altaffilmark{3}, Alice E. Shapley\altaffilmark{4}, Daniel B. Nestor\altaffilmark{4}, Kristin R. Kulas\altaffilmark{5,6}, David R. Law\altaffilmark{7}, Allison L. Strom\altaffilmark{2}, Gwen C. Rudie\altaffilmark{8,9},  Naveen A. Reddy\altaffilmark{10}, Max Pettini\altaffilmark{11}, Nicholas P. Konidaris\altaffilmark{2}, Gregory Mace\altaffilmark{4}, Keith Matthews\altaffilmark{2}, and Ian S. McLean\altaffilmark{4}}

\slugcomment{Accepted for publication in ApJ}

\shorttitle{The \lya\ Properties of Faint Galaxies at $z\sim2$--3}
\shortauthors{Erb et al.}

\altaffiltext{*}{Based on data obtained at the W. M. Keck Observatory, which is operated as a scientific partnership among the California Institute of Technology, the University of California, and the National Aeronautics and Space Administration, and was made possible by the generous financial support of the W. M. Keck Foundation.}

\altaffiltext{1}{Center for Gravitation, Cosmology and Astrophysics, Department of Physics, University of Wisconsin Milwaukee, 1900 E.\ Kenwood Blvd., Milwaukee, WI 53211, USA; \url{erbd@uwm.edu}}
\altaffiltext{2}{Cahill Center for Astrophysics, California Institute of Technology, 1216 E. California Blvd., MS 249-17, Pasadena, CA 91125, USA}
\altaffiltext{3}{Astronomical Observatory, Volgina 7, 11060 Belgrade, Serbia}
\altaffiltext{4}{University of California, Los Angeles, Department of Physics and Astronomy, 430 Portola Plaza, Los Angeles, CA 90095, USA}
\altaffiltext{5}{NASA Ames Research Center, Bldg. 211, Room 112, Moffett Field, CA 94035-1000, USA}
\altaffiltext{6}{NASA Postdoctoral Fellow}
\altaffiltext{7}{Dunlap Institute for Astronomy and Astrophysics, University of Toronto, 50 St. George Street, Toronto, Ontario M5S 3H4, Canada}
\altaffiltext{8}{Carnegie Observatories, 813 Santa Barbara Street, Pasadena, CA 91101, USA}
\altaffiltext{9}{Carnegie-Princeton Fellow}
\altaffiltext{10}{Department of Physics and Astronomy, University of California, Riverside, 900 University Avenue, Riverside, CA 92521, USA}
\altaffiltext{11}{Institute of Astronomy, Madingley Road, Cambridge CB3 0HA, UK}

\begin{abstract}
We study the \lya\ profiles of 36 spectroscopically-detected \lya-emitters (LAEs) at \ztwo--3, using Keck MOSFIRE to measure systemic redshifts and velocity dispersions from rest-frame optical nebular emission lines. The sample has a median optical magnitude ${\cal R}=26.0$, and ranges from ${\cal R}\simeq23$ to ${\cal R}>27$, corresponding to rest-frame UV absolute magnitudes $M_{\rm UV}\simeq-22$ to $M_{\rm UV}>-18.2$. Dynamical masses range from $M_{\rm dyn}<1.3\times10^8$ \msun\  to $M_{\rm dyn}=6.8\times 10^{9}$ \msun, with a median value of $M_{\rm dyn}=6.3\times10^8$ \msun.  Thirty of the 36 \lya\ emission lines are redshifted with respect to the systemic velocity with at least $1\sigma$ significance, and the velocity offset with respect to systemic \dvlya\ is correlated with ${\cal R}$-band magnitude, $M_{\rm UV}$, and the velocity dispersion measured from nebular emission lines with $>3\sigma$ significance:\ brighter galaxies with larger velocity dispersions tend to have larger values of \dvlya. We also make use of a comparison sample of 122 UV-color-selected ${\cal R}<25.5$ galaxies at $z\sim2$, all with \lya\ emission and systemic redshifts measured from nebular emission lines.  Using the combined LAE and comparison samples for a total of 158 individual galaxies, we find that \dvlya\ is anti-correlated with the \lya\ equivalent width with $7\sigma$ significance. Our results are consistent with a scenario in which the \lya\ profile is determined primarily by the properties of the gas near the systemic redshift; in such a scenario, the opacity to \lya\ photons in lower mass galaxies may be reduced if large gaseous disks have not yet developed and if the gas is ionized by the harder spectrum of young, low metallicity stars.
\end{abstract}


\section{Introduction}
\label{sec:intro}

A variety of recent studies have emphasized the importance of faint, low mass galaxies at high redshifts. The steep faint-end slope of the UV luminosity function at $z\gtrsim2$ indicates that faint galaxies make a significant contribution to the global star formation rate density  \citep{rs09,bio+12,asr+14}, and large numbers of low mass galaxies are likely to have been required to reionize the universe \citep{kf12,rfs+13}.  Because the typical galaxy at very high redshift is likely to be young, low in mass and metallicity and relatively unevolved, detailed studies of plausibly similar objects at somewhat lower redshifts may shed light on the physical conditions in galaxies in the early universe (e.g.\ \citealt{eps+10}).

If faint, low mass galaxies have strong \lya\ emission, they can be relatively easily selected by narrowband imaging with a filter tuned to the wavelength of \lya\ at a redshift of interest (e.g.\ \citealt{hm96}). Galaxies with a wide range of masses may be strong \lya-emitters (LAEs), but the average galaxy selected as an LAE is fainter in the continuum and lower in mass than typical galaxies selected via their continuum light in magnitude-limited surveys \citep{kse+10}. Stellar masses of LAEs at $z\sim2$--3 have been determined through spectral energy distribution (SED) fitting, with typical galaxies having masses of 3--10$\times10^8$ \msun\  and little reddening \citep{gvg+06,gap+11,mrm+14,vba+14}. Because of the faintness of these objects, more detailed spectroscopic studies have been difficult.

A wide variety of observational and theoretical studies have addressed the escape of \lya\ emission from galaxies. Because \lya\ is a resonance line, \lya\ photons scatter in both frequency and space, resulting in significant modification of the intrinsic \lya\ profile. The emergent profile depends on many factors, including the geometry, column density and covering fraction of neutral hydrogen, the kinematics of galactic outflows, and the dust content (e.g.\ \citealt{vsm06} and references therein).  The ubiquity of outflows in galaxies at high redshift results in an asymmetric \lya\ profile that is almost always redshifted with respect to the systemic velocity; the simple model invoked to explain this redshift is that of a spherical outflow in which the \lya\ photons that reach an observer are backscattered in the direction of the observer from the far side of the receding outflow, and thus acquire a frequency shift which allows them to pass through the gas in the bulk of the galaxy without additional scattering \citep{pss+01,ssp+03}. Real \lya\ profiles are often more complex than this simple model predicts, and much work has gone into a variety of radiative transfer models aimed at reproducing observed \lya\ profiles (e.g.\ \citealt{vsat08,ksk+12,cbh+13,ldo13,dsol14}, among others).

The ultimate goal of such models is to extract physical galaxy properties such as the outflow velocity and the column density or covering fraction of neutral hydrogen from the observed line profile. In practice, however, the number and complexity of the factors contributing to the line profile often result in degeneracies that make this difficult or impossible. The problem is compounded by the low spectral resolution at which most high redshift observations of \lya\ are made.   In spite of these difficulties, however, some general principles can be stated. An increase in the outflow velocity generally results in an increase in the redshift of \lya\ emission; however, this increase is both nonlinear and non-monotonic, since if the outflow velocity is high enough, \lya\ photons produced in \HII\ regions are out of resonance with the outflowing gas and can escape closer to the systemic velocity \citep{vsm06,vosh14}. An increase in the amount or velocity dispersion of neutral hydrogen at the systemic velocity also results in a larger \lya\ redshift, since the escaping photons must acquire larger frequency shifts in order to escape the neutral gas \citep{vsm06,ses+10}.  \lya\ photons are also preferentially absorbed by dust, since their longer path length due to multiple scatterings increases the likelihood of absorption (\citealt{cf93}, but c.f.\ \citealt{n91}).  In addition, the \lya\ profile is likely to depend on the angle at which the galaxy is observed \citep{lsa09,vdb+12,zw13,son+14b}. In combination, these factors result in a highly complex relationship between the physical conditions in galaxies and the emergent \lya\ profile.

Samples of \lya-selected galaxies at high redshift with measurements of the systemic redshift have so far been small (and the galaxies observed  have tended to be bright), due to the difficulty of obtaining large samples of nebular emission lines in the near-IR \citep{mfr+11,fhg+11,cbh+13,hos+13,gfg+13,mrm+14,rmr+14, sfg+14}. \citet{son+14} have collected previous measurements from the literature and combined them with new data for a sample of 22 LAEs with systemic redshifts; these authors find that the velocity offset of \lya\ emission from the systemic velocity is smaller in these LAEs than in continuum-selected galaxies at the same redshift, and emphasize the role of the covering fraction and column density of \ion{H}{1} in determining \lya\ escape.

In this work we present a study of 36 LAEs at $z\sim2$--3 with systemic redshifts measured from early observations with MOSFIRE, the near-IR multi-object spectrograph recently commissioned on the Keck I telescope. The sensitivity and multiplexing efficiency of MOSFIRE allow large samples of such measurements to be obtained in a relatively short time, and for fainter objects than have been studied previously. Thus our sample is larger than previously published samples of LAEs with systemic redshifts, and includes objects with significantly fainter continuum magnitudes, with ${\cal R}$ ranging from ${\cal R} \simeq 23$ to ${\cal R} > 27$.

We describe the observations and data reduction in Section \ref{sec:obs}, and discuss constraints on the dynamical masses and star formation rates of the sample galaxies in Section \ref{sec:mdyn_sfr}. The \lya\ profiles are presented in Section \ref{sec:profiles}, and in Section \ref{sec:disc} we summarize our results and discuss their implications. We assume the nine-year WMAP cosmological parameters of $H_0=69$ \kms\ Mpc$^{-1}$, $\Omega_{\rm m}=0.29$, and $\Omega_{\Lambda}=0.71$ \citep{wmap09} throughout. 

\begin{deluxetable*}{l l l r r r r}
\tablewidth{0pt}
\tabletypesize{\footnotesize}
\tablecaption{MOSFIRE Observations\label{tab:obs}}
\tablehead{
\colhead{} &
\colhead{} &
\colhead{} &
\multicolumn{2}{c}{$H$-band} &
\multicolumn{2}{c}{$K$-band} \\
\hline
\colhead{Object} & 
\colhead{RA} & 
\colhead{Dec} & 
\colhead{Date Observed} & 
\colhead{Integration Time} & 
\colhead{Date Observed} & 
\colhead{Integration Time} \\
\colhead{} & 
\colhead{(J2000)} & 
\colhead{(J2000)} & 
\colhead{} &
\colhead{(s)} & 
\colhead{} &
\colhead{(s)} 
}
\startdata
Q1700-BNB17\tablenotemark{a} & 17:01:33.685 & 64:06:53.143 & 21 June 2013 & 3578 & ... & ... \\
Q1700-BNB18\tablenotemark{a} & 17:00:50.588 & 64:07:28.636 & 21 June 2013 & 3578 & ... & ... \\
Q1700-BNB19\tablenotemark{a} & 17:01:43.579 & 64:12:52.591 & 13 Sept 2012 & 3578 & 8 May, 5 June 2012 & 8946 \\
Q1700-BNB26\tablenotemark{a} & 16:59:59.103 & 64:06:51.363 & 21 June 2013 & 3578 & ... & ... \\
Q1700-BNB27 & 17:00:46.215 & 64:19:04.119 & 21 June 2013 & 3578 & ... & ... \\
Q1700-BNB29 & 17:01:51.274 & 64:08:24.741 & 21 June 2013 & 3578 & ... & ... \\
Q1700-BNB36 & 17:01:18.709 & 64:12:40.142 & 13 Sept 2012 & 3578 & 8 May, 5 June 2012 & 8946 \\
Q1700-BNB42 & 17:01:31.700 & 64:08:50.079 & 21 June 2013 & 3578 & ... & ... \\
Q1700-BNB47 & 17:00:31.746 & 64:15:38.774 & 21 June 2013 & 3578 & ... & ... \\
Q1700-BNB51 & 17:00:36.169 & 64:07:30.651 & 21 June 2013 & 3578 & ... & ... \\
Q1700-BNB88 & 17:01:27.241 & 64:07:29.465 & 21 June 2013 & 3578 & ... & ... \\
Q1700-BNB93 & 17:00:32.620 & 64:16:28.437 & 21 June 2013 & 3578 & ... & ... \\
Q1700-BNB95 & 17:01:23.458 & 64:14:40.281 & 13 Sept 2012 & 3578 & 8 May, 5 June 2012 & 8946 \\
Q1700-BNB104 & 17:00:51.786 & 64:14:58.858 & ... & ... & 12 Sept 2012 & 5368 \\
Q1700-BNB115 & 17:01:47.623 & 64:13:23.974 & 13 Sept 2012 & 3578 & ... & ... \\
Q1700-BNB153 & 17:01:04.844 & 64:12:52.376 & ... & ... & 8 May, 5 June 2012 & 8946 \\
Q1700-BNB157 & 17:00:21.513 & 64:07:20.169 & 21 June 2013 & 3578 & ... & ... \\
SSA22-001\tablenotemark{a} & 22:17:32.453 & 00:11:33.513 & ... & ... & 13 Sept 2012 & 5368 \\
SSA22-003 & 22:17:24.795 & 00:17:16.998 & ... & ... & 15, 16 Sept 2012 & 13956 \\
SSA22-004\tablenotemark{a} & 22:17:28.034 & 00:14:29.384 & ... & ... & 13 Sept 2012 & 5368 \\
SSA22-006 & 22:17:24.833 & 00:11:16.002 & ... & ... & 29, 30 June 2012 & 8767 \\
SSA22-008\tablenotemark{a} & 22:17:21.126 & 00:15:27.287 & ... & ... & 13 Sept 2012 & 5368 \\
SSA22-009\tablenotemark{a} & 22:17:28.332 & 00:12:11.540 & ... & ... & 29, 30 June 2012 & 8767 \\
SSA22-012\tablenotemark{a} & 22:17:31.714 & 00:16:57.262 & ... & ... & 15, 16 Sept 2012 & 13956 \\
SSA22-013 & 22:17:27.197 & 00:16:21.299 & ... & ... & 29, 30 June 2012 & 8767 \\
SSA22-014 & 22:17:19.274 & 00:14:50.137 & ... & ... & 29, 30 June 2012 & 8767 \\
SSA22-021 & 22:17:18.793 & 00:15:17.315 & ... & ... & 29, 30 June 2012 & 8767 \\
SSA22-042 & 22:17:21.505 & 00:17:04.048 & ... & ... & 29, 30 June 2012 & 8767 \\
SSA22-046 & 22:17:21.487 & 00:14:53.923 & ... & ... & 13 Sept 2012 & 5368 \\
SSA22-062 & 22:17:22.882 & 00:14:41.043 & ... & ... & 29, 30 June 2012 & 8767 \\
SSA22-063 & 22:17:23.335 & 00:15:52.333 & ... & ... & 29, 30 June 2012 & 8767 \\
SSA22-066 & 22:17:20.877 & 00:15:11.123 & ... & ... & 29, 30 June 2012 & 8767 \\
SSA22-067 & 22:17:36.306 & 00:13:11.416 & ... & ... & 13 Sept 2012 & 5368 \\
SSA22-072 & 22:17:31.259 & 00:17:31.810 & ... & ... & 15, 16 Sept 2012 & 13956 \\
SSA22-078 & 22:17:37.704 & 00:16:47.841 & ... & ... & 13 Sept 2012 & 5368 \\
SSA22-082 & 22:17:35.471 & 00:16:47.238 & ... & ... & 13 Sept 2012 & 5368 
\enddata
\tablenotetext{a}{These 9 LAEs also satisfy the UV color selection criteria for $z\sim2$ BX galaxies \citep{ssp+04} or $z\sim3$ Lyman
break galaxies \citep{sas+03}: Q1700-BNB17 = BX239, Q1700-BNB18 = BX313, Q1700-BNB19 = BX754, Q1700-BNB26 = BX235, 
SSA22-001 = D3, SSA22-004 = MD23, SSA22-008 = C28, SSA22-009 = C9, and SSA22-012 = M28.}
\end{deluxetable*}

\begin{deluxetable*}{l r r r r r r r r}
\tablewidth{0pt}
\tabletypesize{\footnotesize}
\tablecaption{\lya\ Properties\label{tab:lya}}
\tablehead{
\colhead{Object} & 
\colhead{$\cal R$\tablenotemark{a}} & 
\colhead{$z_{\rm neb}$\tablenotemark{b}} & 
\colhead{$z_{\lya}$} & 
\colhead{$F_{\lya}$\tablenotemark{c}} &
\colhead{$\Delta v_{\lya}$\tablenotemark{d}} &
\colhead{Spec $W_{\lya}$\tablenotemark{e}} &
\colhead{$A_{\lya}$\tablenotemark{f}} &
\colhead{Phot $W_{\lya}$\tablenotemark{g}} \\
\colhead{} & 
\colhead{} & 
\colhead{} & 
\colhead{} & 
\colhead{($\times 10^{-17}$)} & 
\colhead{(\kms)} & 
\colhead{(\AA)} &
\colhead{} &
\colhead{(\AA)} 
}
\startdata
Q1700-BNB17 & $24.48$ & 2.2838 & 2.293 & $0.62 \pm 0.13$ & $840\pm69$ & $>\phantom{0}9$ & $0.4\pm0.3$ & $33$ \\
Q1700-BNB18 & $24.75$ & 2.3135 & 2.317 & $1.59 \pm 0.19$ & $344\pm43$ & $>11$ & $0.1\pm0.1$ & $30$ \\
Q1700-BNB19 & $23.94$ & 2.2837 & 2.288 & $11.26 \pm 1.91$ & $393\pm127$ & $>21$ & $0.4\pm0.1$ & $35$ \\
Q1700-BNB26 & $25.24$ & 2.2776 & 2.280 & $2.11 \pm 0.21$ & $210\pm44$ & $>13$ & $0.4\pm0.1$ & $34$ \\
Q1700-BNB27 & $23.09$ & 2.2928 & 2.295 & $2.67 \pm 0.17$ & $237\pm17$ & $>18$ & $0.4\pm0.1$ & $35$ \\
Q1700-BNB29 & $25.07$ & 2.2930 & 2.298 & $3.39 \pm 0.46$ & $455\pm49$ & $11\pm2$ & $0.0\pm0.1$ & $28$ \\
Q1700-BNB36 & $24.84$ & 2.2949 & 2.298 & $3.60 \pm 0.25$ & $264\pm24$ & $>39$ & $0.3\pm0.1$ & $47$ \\
Q1700-BNB42 & $26.01$ & 2.2742 & 2.275 & $5.44 \pm 0.25$ & $\phantom{0}73\pm22$ & $>60$ & $0.1\pm0.0$ & $102$ \\
Q1700-BNB47 & $26.07$ & 2.2935 & 2.296 & $1.85 \pm 0.27$ & $273\pm61$ & $>13$ & $0.2\pm0.2$ & $33$ \\
Q1700-BNB51 & $26.11$ & 2.3100 & 2.313 & $2.79 \pm 0.28$ & $245\pm37$ & $>23$ & $0.5\pm0.1$ & $91$ \\
Q1700-BNB88 & $26.14$ & 2.3002 & 2.301 & $1.14 \pm 0.17$ & $\phantom{0}45\pm64$ & $>16$ & $1.1\pm0.1$ & $43$ \\
Q1700-BNB93 & $>26.85$ & 2.3254 & 2.327 & $2.48 \pm 0.21$ & $144\pm20$ & $>21$ & $1.7\pm0.6$ & $73$ \\
Q1700-BNB95 & $25.71$ & 2.3064 & 2.312 & $3.85 \pm 1.12$ & $508\pm258$ & $>10$ & $0.5\pm0.4$ & $37$ \\
Q1700-BNB104 & $>26.48$ & 2.2942 & 2.295 & $1.73 \pm 0.20$ & $\phantom{0}27\pm56$ & $>20$ & $1.4\pm0.3$ & $106$ \\
Q1700-BNB115 & $26.39$ & 2.3064 & 2.308 & $1.19 \pm 0.11$ & $145\pm60$ & $17\pm2$ & $0.8\pm0.2$ & $31$ \\
Q1700-BNB153 & $>26.85$ & 2.2903 & 2.291 & $1.86 \pm 0.19$ & $\phantom{0}64\pm47$ & $>26$ & $1.2\pm0.3$ & $85$ \\
Q1700-BNB157 & $>26.72$ & 2.3139 & 2.317 & $2.20 \pm 0.40$ & $299\pm94$ & $>16$ & $0.0\pm0.1$ & $203$ \\
SSA22-001 & $23.92$ & 3.0690 & 3.075 & $1.02 \pm 0.06$ & $442\pm19$ & $12\pm1$ & $0.1\pm0.0$ & $21$ \\
SSA22-003 & $24.42$ & 3.0965 & 3.097 & $1.17 \pm 0.04$ & $\phantom{0}37\pm16$ & $40\pm4$ & $1.0\pm0.1$ & $39$ \\
SSA22-004 & $24.34$ & 3.0788 & 3.092 & $1.64 \pm 0.12$ & $970\pm32$ & $\phantom{0}4\pm2$ & $-0.6\pm0.2$ & $36$ \\
SSA22-006 & $>27.00$ & 3.0691 & 3.076 & $0.40 \pm 0.04$ & $508\pm68$ & $>15$ & $0.1\pm0.1$ & $90$ \\
SSA22-008 & $24.87$ & 3.0692 & 3.076 & $2.38 \pm 0.14$ & $501\pm25$ & $31\pm4$ & $0.0\pm0.0$ & $30$ \\
SSA22-009 & $25.84$ & 3.0688 & 3.071 & $1.71 \pm 0.06$ & $162\pm12$ & $>46$ & $0.4\pm0.0$ & $87$ \\
SSA22-012 & $24.75$ & 3.0902 & 3.094 & $0.74 \pm 0.09$ & $278\pm50$ & $>15$ & $0.3\pm0.1$ & $31$ \\
SSA22-013 & $25.98$ & 3.0919 & 3.095 & $0.79 \pm 0.05$ & $227\pm$33 & $>32$ & $0.3\pm0.0$ & $122$ \\
SSA22-014 & $25.82$ & 3.0631 & 3.067 & $0.69 \pm 0.09$ & $288\pm32$ & $>\phantom{0}8$ & $0.0\pm0.1$ & $62$ \\
SSA22-021 & $>27.00$ & 3.0670 & 3.070 & $0.91 \pm 0.05$ & $221\pm19$ & $>46$ & $0.4\pm0.0$ & $92$ \\
SSA22-042 & $25.50$ & 3.0666 & 3.072 & $0.34 \pm 0.04$ & $398\pm49$ & $>16$ & $0.2\pm0.1$ & $35$ \\
SSA22-046 & $>27.00$ & 3.0975 & 3.100 & $1.02 \pm 0.11$ & $183\pm27$ & $>10$ & $0.2\pm0.1$ & $>277$ \\
SSA22-062 & $26.53$ & 3.0551 & 3.057 & $0.94 \pm 0.04$ & $140\pm15$ & $>36$ & $0.5\pm0.1$ & $49$ \\
SSA22-063 & $26.55$ & 3.0978 & 3.099 & $0.53 \pm 0.03$ & $\phantom{0}88\pm32$ & $>27$ & $0.8\pm0.1$ & $83$ \\
SSA22-066 & $26.64$ & 3.0645 & 3.066 & $0.19 \pm 0.05$ & $111\pm194$ & $>\phantom{0}6$ & $-0.1\pm4.2$ & $20$ \\
SSA22-067 & $26.40$ & 3.1013 & 3.106 & $0.60 \pm 0.07$ & $344\pm54$ & $>17$ & $0.2\pm0.1$ & $40$ \\
SSA22-072 & $27.00$ & 3.0845 & 3.084 & $0.44 \pm 0.13$ & $-37\pm242$ & $>11$ & $1.7\pm0.9$ & $102$ \\
SSA22-078 & $25.95$ & 3.0870 & 3.090 & $0.18 \pm 0.06$ & $220\pm258$ & $>\phantom{0}5$ & $0.4\pm0.2$ & $46$ \\
SSA22-082 & $>27.00$ & 3.0873 & 3.087 & $0.56 \pm 0.08$  & $-22\pm86$ & $>12$ & $1.4\pm0.5$ & $72$ 
\enddata
\tablenotetext{a}{${\cal R}$-band magnitude. 3$\sigma$ limits are given for non-detections.}
\tablenotetext{b}{Systemic redshift from nebular emission lines. Redshifts are measured from \OIII$\lambda$5007 emission, except in the cases of Q1700-BNB104 and Q1700-BNB153,
for which \Ha\ emission is used.}
\tablenotetext{c}{Line fluxes are given in units of $10^{-17}$ erg s$^{-1}$
  cm$^{-2}$.}
\tablenotetext{d}{Velocity offset of \lya\ and nebular emission as defined in Equation \ref{eq:dvlya}; 
positive velocities indicate that \lya\ emission is redshifted with respect to nebular emission.}
\tablenotetext{e}{Spectroscopically measured rest-frame
  \lya\ equivalent width.}
\tablenotetext{f}{Ratio of spectroscopic \lya\ equivalent width
  blueward of systemic velocity to spectroscopic \lya\ equivalent width
  redward of systemic velocity.}
\tablenotetext{g}{Rest-frame \lya\ equivalent width from narrow-band
  imaging. Values for SSA22 objects from \citet{nsss11,nsk+13}.}
\end{deluxetable*}

\begin{deluxetable*}{l r r r r r r r r}
\tablewidth{0pt}
\tabletypesize{\footnotesize}
\tablecaption{Emission Line Measurements from MOSFIRE Spectra\label{tab:mosfire}}
\tablehead{
\colhead{Object} & 
\colhead{$\cal R$\tablenotemark{a}} & 
\colhead{$z_{\rm neb}$\tablenotemark{b}} & 
\colhead{$\sigma$\tablenotemark{b}} & 
\colhead{$F_{\rm H\beta}$\tablenotemark{c}} &
\colhead{$F_{\rm[OIII]\lambda 4959}$\tablenotemark{c}} &
\colhead{$F_{\rm[OIII]\lambda 5007}$\tablenotemark{c}} &
\colhead{$F_{\rm H\alpha}$\tablenotemark{c}} &
\colhead{$F_{\rm[NII]\lambda 6564}$\tablenotemark{c}} \\
\colhead{} & 
\colhead{(AB)} & 
\colhead{} & 
\colhead{(\kms)} & 
\colhead{($\times 10^{-17}$)} &
\colhead{($\times 10^{-17}$)} &
\colhead{($\times 10^{-17}$)} &
\colhead{($\times 10^{-17}$)} &
\colhead{($\times 10^{-17}$)} 
}
\startdata
Q1700-BNB17 & 24.48 & 2.2838 & $59^{+11}_{-12}$ &	$<2.06$ & $1.14\pm0.19$ & $3.91\pm0.95$ & ... & ... \\
Q1700-BNB18 & 24.75 & 2.3135 & $60^{+6}_{-6}$ &	$<0.50$ & $<1.74$ & $3.22\pm0.15$ & ... & ... \\
Q1700-BNB19 & 23.94 & 2.2837 & $101^{+4}_{-4}$ &	$1.25\pm0.08$ & $1.07\pm0.05$ & $3.68\pm0.22$ & $3.51\pm0.12$ & $0.65\pm0.07$ \\
Q1700-BNB26 & 25.24 & 2.2776 & $79^{+7}_{-7}$ &	$0.92\pm0.20$ & $1.84\pm0.30$ & $4.54\pm0.53$ & ... & ... \\
Q1700-BNB27 & 23.09 & 2.2928 & $<29$ &	$<0.39$ & $0.41\pm0.13$ & $1.29\pm0.12$ & ... & ... \\
Q1700-BNB29 & 25.07 & 2.2930 & $<42$ &	$<0.47$ & $<0.51$ & $0.65\pm0.15$ & ... & ... \\
Q1700-BNB36 & 24.84 & 2.2949 & $58^{+6}_{-6}$ &	$0.55\pm0.09$ & $<0.64$ & $3.04\pm0.41$ & $1.36\pm0.09$ & $<0.12$ \\
Q1700-BNB42 & 26.01 & 2.2742 & $37^{+15}_{-20}$ &	$<0.79$ & $<2.27$ & $1.56\pm0.28$ & ... & ... \\
Q1700-BNB47 & 26.07 & 2.2935 & $36^{+8}_{-10}$ &	$0.48\pm0.16$ & $<0.95$ & $1.95\pm0.35$ & ... & ... \\
Q1700-BNB51 & 26.11 & 2.3100 & $40^{+7}_{-8}$ &	$<0.46$ & $<0.95$ & $1.52\pm0.12$ & ... & ... \\
Q1700-BNB88 & 26.14 & 2.3002 & $<41$ &	$<0.37$ & $<0.62$ & $0.61\pm0.21$ & ... & ... \\
Q1700-BNB93 & $>26.85$ & 2.3254 & $<44$ &	$<0.43$ & $<0.71$ & $0.79\pm0.16$ & ... & ... \\
Q1700-BNB95 & 25.71 & 2.3064 & $34^{+4}_{-4}$ &	$<0.68$ & $1.06\pm0.07$ & $2.69\pm0.13$ & $1.23\pm0.11$ & $<0.12$ \\
Q1700-BNB104 & $>26.48$ & 2.2942 & $29^{+12}_{-17}$ &	... & ... & ... & $0.89\pm0.15$ & $<0.27$ \\
Q1700-BNB115 & 26.39 & 2.3064 & $<35$ &	$1.47\pm0.34$ & $0.20\pm0.05$ & $0.67\pm0.10$ & ... & ... \\
Q1700-BNB153 & $>26.85$ & 2.2903 & $<39$ &	... & ... & ... & $0.49\pm0.05$ & $<0.20$ \\
Q1700-BNB157 & $>26.72$ & 2.3139 & $32^{+14}_{-20}$ &	$<0.47$ & $<1.73$ & $0.85\pm0.12$ & ... & ... \\
SSA22-001 & 23.92 & 3.0690 & $112^{+ 4}_{- 4}$ & $1.40\pm0.12$ & $1.35\pm0.28$ & $5.80\pm0.19$ & ... & ... \\
SSA22-003 & 24.42 & 3.0965 & $ 54^{+ 6}_{- 6}$ & $0.37\pm0.07$ & $0.55\pm0.08$ & $1.73\pm0.13$ & ... & ... \\
SSA22-004 & 24.34 & 3.0788 & $ 83^{+ 3}_{- 3}$ & $1.20\pm0.09$ & $1.60\pm0.09$ & $4.81\pm0.13$ & ... & ... \\
SSA22-006 & $>27.00$ & 3.0691 & $173^{+ 9}_{- 9}$ & $0.85\pm0.10$ & $0.65\pm0.21$ & $3.32\pm0.15$ & ... & ... \\
SSA22-008 & 24.87 & 3.0692 & $ 82^{+ 1}_{- 1}$ & $1.11\pm0.05$ & $2.31\pm0.13$ & $7.35\pm0.09$ & ... & ... \\
SSA22-009 & 25.84 & 3.0687 & $ 40^{+ 4}_{- 4}$ & $0.16\pm0.04$ & $0.36\pm0.09$ & $1.38\pm0.07$ & ... & ... \\
SSA22-012 & 24.75 & 3.0902 & $ 96^{+15}_{-15}$ & $0.57\pm0.10$ & $0.42\pm0.11$ & $1.06\pm0.14$ & ... & ... \\
SSA22-013 & 25.98 & 3.0919 & $ 57^{+16}_{-18}$ & $<0.26$ & $<0.24$ & $0.86\pm0.15$ & ... & ... \\
SSA22-014 & 25.82 & 3.0631 & $ 34^{+ 9}_{-10}$ & $<0.32$ & $<0.31$ & $1.30\pm0.15$ & ... & ... \\
SSA22-021 & $>27.00$ & 3.0670 & $ 59^{+10}_{-11}$ & $<0.59$ & $<0.58$ & $0.99\pm0.13$ & ... & ... \\
SSA22-042 & 25.50 & 3.0666 & $ 76^{+ 5}_{- 6}$ & $<0.43$ & $0.64\pm0.12$ & $1.30\pm0.07$ & ... & ... \\
SSA22-046 & $>27.00$ & 3.0975 & $ 27^{+11}_{-16}$ & $<0.25$ & $0.29\pm0.06$ & $0.77\pm0.13$ & ... & ... \\
SSA22-062 & 26.53 & 3.0551 & $ 30^{+ 7}_{- 8}$ & $<0.10$ & $<0.24$ & $0.61\pm0.06$ & ... & ... \\
SSA22-063 & 26.55 & 3.0978 & $ 40^{+16}_{-20}$ & $<0.24$ & $<0.17$ & $0.46\pm0.12$ & ... & ... \\
SSA22-066 & 26.64 & 3.0645 & $ 58^{+12}_{-13}$ & $<0.27$ & $<0.39$ & $0.79\pm0.11$ & ... & ... \\
SSA22-067 & 26.40 & 3.1013 & $ 65^{+18}_{-19}$ & $<0.23$ & $<0.83$ & $1.18\pm0.28$ & ... & ... \\
SSA22-072 & 27.00 & 3.0845 & $ 34^{+12}_{-15}$ & $<0.23$ & $0.34\pm0.11$ & $0.86\pm0.16$ & ... & ... \\
SSA22-078 & 25.95 & 3.0870 & $<35$ & $<0.12$ & $<1.50$ & $0.24\pm0.05$ & ... & ... \\
SSA22-082 & $>27.00$ & 3.0873 & $ 22^{+ 6}_{- 7}$ & $<0.10$ & $<0.81$ & $0.91\pm0.07$ & ... & ...
\enddata
\tablenotetext{a}{${\cal R}$-band magnitude. 3$\sigma$ limits are given for non-detections.}
\tablenotetext{b}{Redshifts and line widths are measured from \OIII$\lambda$5007 emission, except in the cases of Q1700-BNB104 and Q1700-BNB153, 
for which \Ha\ emission is used. Velocity dispersions have been corrected for instrumental broadening.}
\tablenotetext{c}{Line fluxes are given in units of $10^{-17}$ erg s$^{-1}$
  cm$^{-2}$.}
\end{deluxetable*}

\section{Observations and Data Reduction}
\label{sec:obs}

The sample of 36 LAEs is comprised of 17 objects at $z\approx2.3$ and 19 at $z\approx3.1$. The \ztwo\ sample is part of the $z=2.3$ protocluster identified in the Q1700 field by \citet{sas+05}, and the $z\sim3$ sample lies in the SSA22 field, which also contains a known overdensity of LAEs at $z=3.09$ \citep{sas+00,myh+05}.

\subsection{Optical imaging and spectroscopy}
\label{sec:opt}
Broadband optical ($U_n G \cal{R}$) imaging in the Q1700 field is described by \citet{sse+05}. Candidate LAEs were selected from a deep narrowband image obtained with the LFC Wide-Field Imager on the Hale 200-inch telescope at Palomar Observatory in July 2007. Observations were conducted under clear conditions with typical seeing of $\sim$1\secpoint2. The LFC camera is an array of six $2048 \times 4096$ pixel back-side illuminated SITe CCDs, covering a field of view that is approximately 24\arcmin\space in diameter. The plate scale is 0\secpoint18/pixel, providing an image scale of 0\secpoint36/pixel in the $2 \times 2$ binned readout mode, which was used for our observations.

The LFC camera was equipped with a custom narrow-band filter centered at $\lambda_{\rm eff}=4010$ \AA\  and with a width of $\textrm{FWHM}=90$ \AA. The data were reduced according to standard procedures in IRAF using the Mosaic Data Reduction Package (MSCRED), and the final image represents an effective exposure time of $22.3$ hrs. The final image reaches a photometric depth of  $NB=26.8$ (AB) for a $3\sigma$ detection in a $3\arcsec$ diameter aperture, equivalent to $N B\sim 27.8$ mag arcsec$^{-2}$.

The selection of objects with an excess in the narrow-band filter also requires an estimate of the continuum at 4010 \AA.  The peak transmission of the narrow-band filter is located between the transmission bands of the $U_n$ and $G$ filters, and we therefore use a linear combination of the two broad-band measurements to predict the narrowband flux in the absence of emission lines,
 \begin{equation}
   UG(4010\,\mbox{\AA}) = 0.63 \times U_n + 0.37 \times G.
 \label{eqn:4010}
 \end{equation}
The coefficients in this equation are derived from the effective wavelengths of the $U_n$ and $G$ filters.  Using this formula, we create a ``$UG$ continuum'' image, which provides an estimate for the spatial extent and shape of the continuum flux at 4010 \AA.
 
Photometry on the narrowband and continuum images was performed with SExtractor, and resulted in a catalog of $\sim6700$ narrowband-detected sources to a limiting magnitude $NB  = 25.5$. \lya-emitter candidates were then selected according to the criterion
 \begin{equation}
   (NB-UG)_{\rm corr} \ge 0.75,
 \label{eq:nb_crit}
 \end{equation}
where the $NB-UG$ color has been corrected by an empirically-defined function $f(U_n-G)$ such that
\begin{equation}
   (NB-UG)_{\rm corr} =  (NB-UG) - f(U_n-G).
 \end{equation}
The function $f(U_n-G)$ is a polynomial fit to the locus of points in the $(NB-UG)$--$(U_n-G)$ color-color diagram, and ranges in value from $-0.06$ to 0.11 mag over the $U_n-G$ range of interest. When this correction is applied, both the average and median of  $(NB-UG)_{\rm corr}$ are near zero after clipping of outliers. This correction accounts for changes in the $U_n$ magnitude with redshift due to increasing blanketing of the $U_n$ passband by the \lya\ forest for galaxies in the redshift range $2<z<2.5$.  The cut given by Equation \ref{eq:nb_crit} then results in a sample of 119 LAE candidates brighter than $NB=25.5$.

Optical follow-up spectroscopy of the LAEs was obtained with the Low Resolution Imaging Spectrometer (LRIS, \citealt{occ+95,ssp+04}) on the Keck I telescope during several observing runs between September 2006 and September 2009. Objects discussed in this paper were observed with the 400/3400 grism and d680 dichroic, providing a spectroscopic resolution of FWHM~$=6.8$ \AA;  for \lya\ observed at $z \sim 2.3$, this corresponds to a velocity resolution of $ \sim 510$ \kms.  Observations were conducted and spectra reduced as described by \citet{ssp+04}. All of the optical observations are described in more detail by \citet{b10}.

Optical imaging and spectroscopy in the SSA22 field has been described extensively elsewhere. The LAEs discussed in this paper are drawn from  the narrowband imaging and spectroscopy discussed by \citet{sas+00} and \citet{nsss11,nsk+13}, and we refer the reader to those papers for details of the observations and data reduction. Spectroscopy of the LAEs was again obtained with LRIS, and most of the SSA22 sample was observed with the 300-line grism, offering a velocity resolution of $\sim530$ \kms\ for \lya\ observed at $z=3.1$ (two of the SSA22 objects were observed with the 600-line grism, which provides a resolution of $\sim220$ \kms\ for \lya\ at $z=3.1$).  

\begin{figure*}[htbp]
\plotone{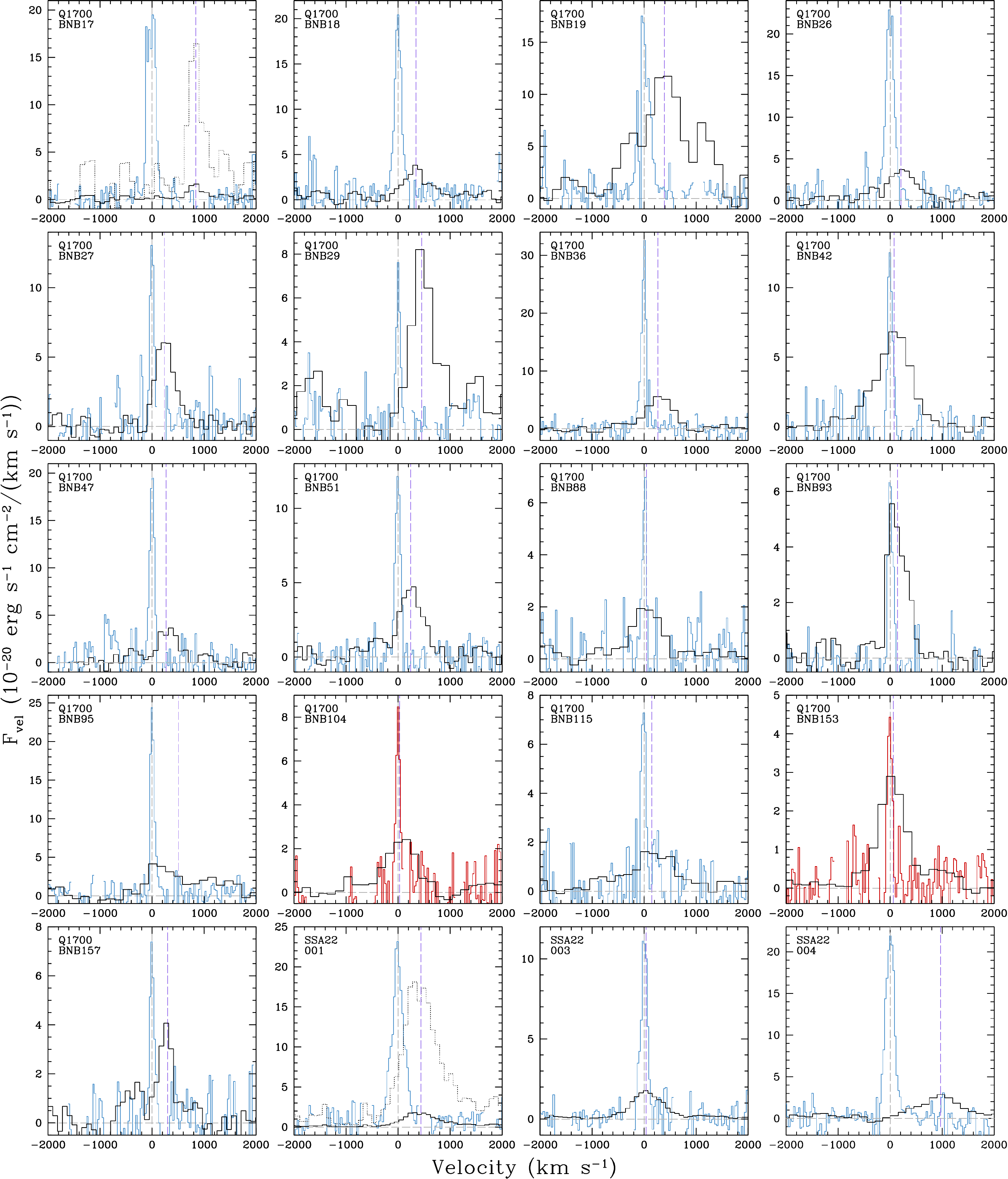}
\caption{\lya\ and nebular emission line profiles of the NB-selected sample. \lya\ profiles are shown in black, \OIII\ $\lambda5007$ profiles in blue, and \Ha\ profiles (Q1700-BNB104 and Q1700-BNB153) in red. All lines are plotted on the same flux scale, in $F_{\rm vel}$ units of flux per unit velocity such that the integral under each emission line is equal to the total flux. For the objects with the weakest \lya\ emission relative to the nebular lines (Q1700-BNB17, SSA22-001, SSA22-006, SSA22-042, SSA22-066, SSA22-072, SSA22-078 and SSA22-082), we also plot the \lya\ profile multiplied by a factor of 10 in dotted lines for clarity. The dashed grey vertical line in each panel indicates zero velocity corresponding to the systemic redshift measured by MOSFIRE, and the dashed purple vertical line indicates the \lya\ velocity offset. The strongest sky line residuals in the MOSFIRE spectra have been masked.}
\label{fig:lyaprofiles}
\end{figure*}

\begin{figure*}[htbp]
\plotone{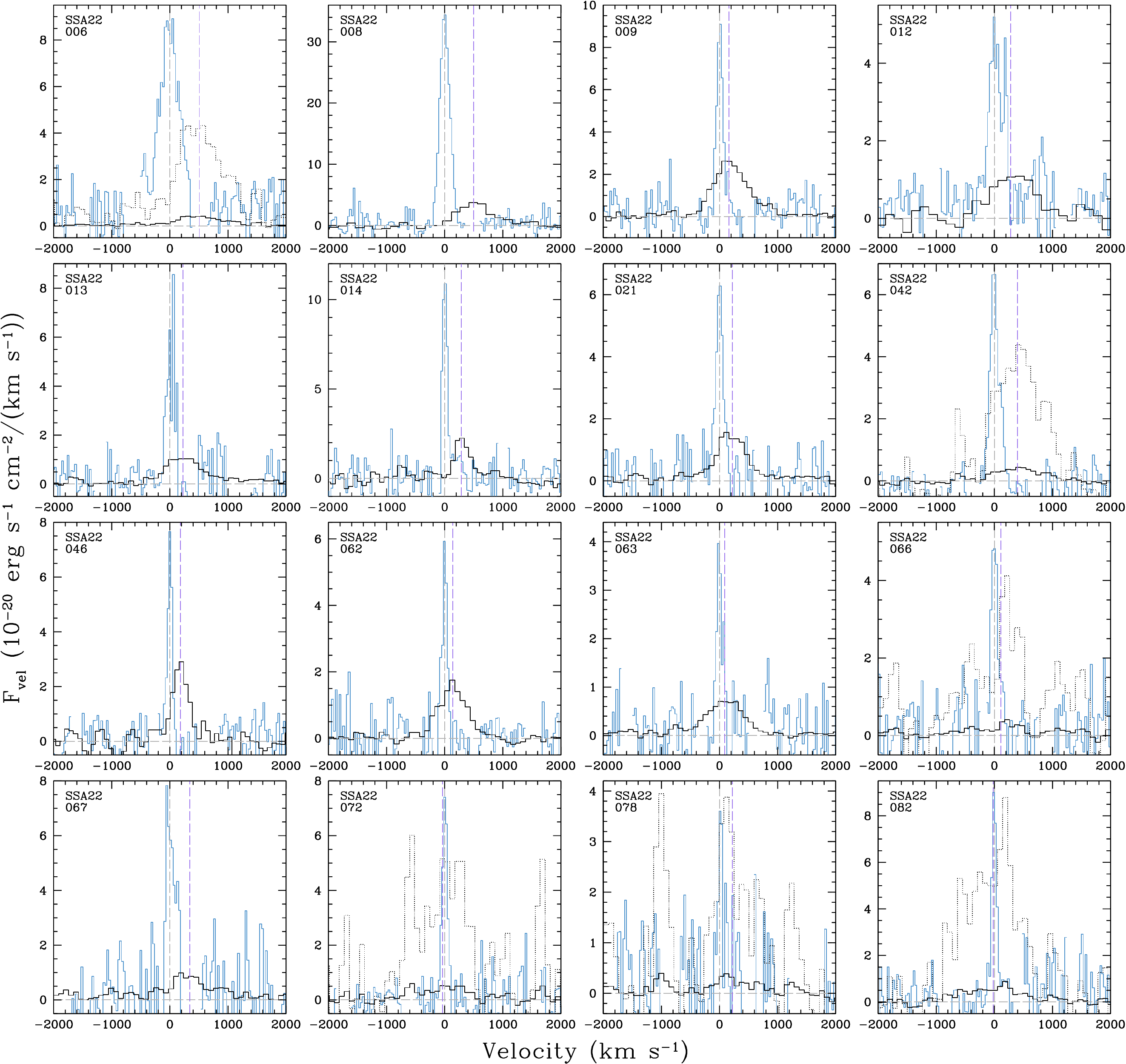}
\figurenum{1}
\caption{Continued.}
\end{figure*}

\subsection{Near-IR spectroscopy}

Near-IR spectra of the rest-frame optical emission lines of the LAE sample were obtained with the  Multi-Object Spectrometer for InfraRed Exploration (MOSFIRE; \citealt{mosfire,mosfire2}) on the Keck I telescope. MOSFIRE uses a configurable slit mechanism to obtain spectra of up to 46 objects simultaneously, over a 6\minpoint1 $\times$ 6\minpoint1 field of view. With the 0\secpoint7 slit width used for this program, MOSFIRE achieves spectral resolution of  $R = 3690$ in the $K$-band and $R=3620$ in $H$.  Some data were obtained during MOSFIRE commissioning science verification in 2012 June, while the bulk of the sample was observed in 2012 September or 2013 June.  Most observations focused on \OIII\ $\lambda5007$ emission, and so were obtained in the $H$-band for the Q1700 LAEs at $z\approx2.3$ and in the $K$-band for the SSA22 LAEs at $z\approx3.1$. Typical exposure times were 1 hour per mask (divided into 15 sets of two dithered 120 s exposures in the $H$-band and 10 sets of two dithered 180 s exposures in the $K$-band), although some objects were observed for longer. Total integration times and dates of observations are given in Table \ref{tab:obs}.

Data were reduced using the publicly available MOSFIRE data reduction pipeline,\footnote{Available at \url{http://code.google.com/p/mosfire/}} which produces combined, wavelength-calibrated and rectified two-dimensional spectra for each slit. Wavelength calibration uses the night sky lines, with typical residual RMS of 0.08 \AA\ in $K$ and 0.06 \AA\ in $H$, corresponding to $\simeq1.1$ \kms. One-dimensional spectra were extracted using MOSPEC, an IDL-based tool developed for the analysis of MOSFIRE spectra (A.\ Strom et al.\ in prep).  In addition to extraction of the 1d spectrum and its associated 1$\sigma$ error spectrum, MOSPEC flux-calibrates the data based on observations of A0V stars, determines a low-order polynomial fit to the continuum, and then performs a simultaneous Gaussian fit to a specified set of emission lines, constraining the lines to have the same redshift and width in a given observation. MOSPEC then reports the redshift, line width, fluxes of all lines measured, and the uncertainty of each. MOSFIRE, the observational procedures, and the data reduction and analysis are described in more detail by \citet{srs+14}. Both the \lya\ profiles from LRIS and the nebular emission line spectra from MOSFIRE are shown in Figure \ref{fig:lyaprofiles}, and the \lya\ and nebular emission line properties of the LAE sample are given in Tables \ref{tab:lya} and \ref{tab:mosfire} respectively. All objects in the sample have at least 3$\sigma$ detections of both \lya\ and \OIII\ $\lambda5007$ or \Ha\ emission lines; most of the spectroscopic measurements have much higher significance, but the sample also includes objects with noisy spectra such as SSA22-078, which has the lowest \lya\ and \OIII\ fluxes in the sample.

\subsection{Comparison sample}

We also make use of a comparison sample of 122 rest-frame UV color-selected galaxies with ${\cal R}<25.5$. These star-forming, \lya-emitting galaxies provide greater dynamic range in \lya\ equivalent width in order to investigate possible correlations. The galaxies in the comparison sample are drawn from the $z\sim2$ spectroscopic survey described by  \citet{ssp+04}. Approximately 40\% of spectroscopically confirmed galaxies in this survey have \lya\ emission, and $\sim10$\% have equivalent widths \wlya$~>20$ \AA; for Lyman break galaxies at $z\sim3$, these fractions are higher at 65\% and 23\% respectively \citep{rsp+08}.  

All of the galaxies in the comparison sample have both \lya\ in emission as spectroscopically observed with LRIS, and systemic redshifts obtained from measurements of nebular emission lines.  94 of the galaxies in the comparison sample were observed with MOSFIRE, and 28 have systemic redshifts from NIRSPEC as previously described by \citet{ses+10} and \citet{ess+06mass}.  Color selection via the BX/BM criteria and spectroscopic observations with LRIS are described by \citet{ssp+04}; 96 of the 122 comparison galaxies were observed with the 400-line grism, 25 with the 600-line grism, and one with the 300-line grism. \citet{srs+14} discuss the initial results obtained from the MOSFIRE spectra of this sample. In this work we use the MOSFIRE measurements only to obtain systemic redshifts for study of the \lya\ profiles; these data will be presented in full elsewhere (A.\ Strom et al.\ in prep).

\section{Dynamical Masses, Star Formation Rates and Line Ratios}
\label{sec:mdyn_sfr}

We would like to estimate the stellar masses and star formation rates of the LAE sample, but because most of the galaxies are extremely faint we have limited ability to infer their stellar population properties. These galaxies are typically not detected in images at $K$-band or longer wavelengths, and we have therefore not attempted SED modeling of their stellar populations. Lack of stellar population information also precludes a test of any potential evolution between the $z\sim3$ and $z\sim2$ LAE samples.
However, half of the sample is covered by {\it HST} imaging, enabling a calculation of dynamical masses from the widths of nebular lines in the MOSFIRE spectra and the sizes. We also calculate lower limits on the star formation rates from the dust-uncorrected rest-frame UV luminosities, and compare the ratios of \lya\ and the rest-frame optical emission lines.

\subsection{Dynamical masses}

\begin{figure*}[htbp]
\plotone{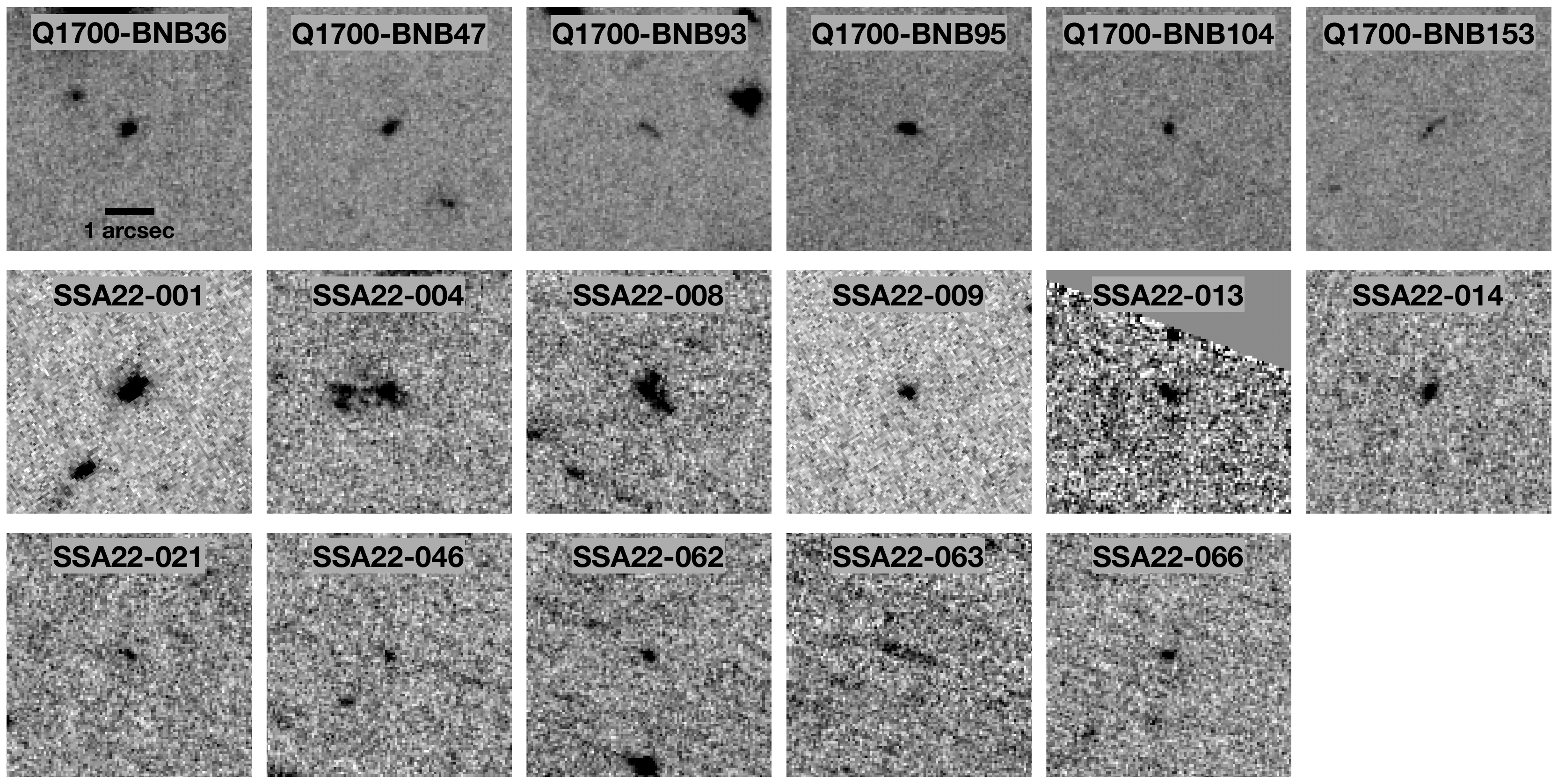}
\caption{{\it HST} ACS F814W images of 17 galaxies in the LAE sample. The scale bar in the upper left image indicates 1\arcsec.}
\label{fig:hst}
\end{figure*}

Nineteen of the 36 galaxies in the LAE sample are covered by {\it HST} ACS and/or WFC3 imaging: six LAEs in the Q1700 field were imaged by  ACS in the course of our Cycle 14 program focusing on the Q1700 protocluster (PI Shapley, data described by \citealt{psl+07}), and images of 13 SSA22 LAEs have been retrieved from the archive. All 19 LAEs are covered by ACS using the F814W filter, and several of the SSA22 galaxies are covered by ACS F625W and WFC3 F160W as well. We measure magnitudes for the LAEs, and determine their sizes by fitting a 2-dimensional \citet{sersic63} profile with GALFIT \citep{phir10}, for which the size $a$ is the effective half-light radius along the semimajor axis. These procedures are described in detail by \citet{lss+12}. Because of the faintness and small sizes of most of these objects, we do not attempt further morphological analysis. Seventeen of the 19 LAEs are significantly detected in the F814W band. We show images of these 17 galaxies in Figure \ref{fig:hst}, and report their magnitudes and sizes in Table \ref{tab:hst}. Including the two non-detections, which are the faintest in the sample, the median F814W magnitude of the 19 LAEs is 26.51.

We calculate dynamical masses using sizes measured in the F814W band, according to the formula
\begin{equation}
M_{\rm dyn} = C \frac{\sigma^2 a}{G},
\label{eq:mdyn}
\end{equation}
where $\sigma$ is the velocity dispersion measured from the width of the nebular emission lines (usually \OIII $\lambda5007$), $a$ is the effective half-light radius described above, and the constant $C$ depends on the mass distribution. Lacking any information about the galaxies' density profiles, we adopt $C=2$, appropriate to a singular isothermal sphere \citep{binneytremaine}, but the value of $C$ remains a source of systematic uncertainty. Many previous studies have assumed $C\approx 2$--3 (see discussion by \citealt{mvd+13}), while others have adopted higher values of $C\approx4$--5 \citep{pss+01,ess+03,rmr+14}; thus the systematic uncertainty due to the unknown value of $C$ is a factor of $\sim2$.

Velocity dispersions (see Table \ref{tab:mosfire}) for the full LAE sample range from 22 \kms\ to 173 \kms\ with a median of 54 \kms; seven of the nebular emission lines are unresolved, and we treat upper limits in velocity dispersion as detections when calculating the median.  The 17 galaxies with F814W detections have a median size of 0.6 kpc. Six of the 17 are unresolved, in which case we report the size of the PSF as an upper limit, and report an upper limit on the dynamical mass (upper limits on dynamical mass are also given when we report an upper limit on the velocity dispersion). Dynamical masses calculated from Equation \ref{eq:mdyn} are given in Table \ref{tab:hst}, and range from $M_{\rm dyn} < 1.3 \times 10^8$ \msun\  to $M_{\rm dyn} = 6.8 \times 10^{9}$ \msun, with a median value of $M_{\rm dyn} = 6.3 \times 10^8$ \msun.  We have again treated upper limits as detections when calculating the median, and we note that uncertainties due to size measurements of very faint objects can be substantial.

CO measurements at $z\sim2$ indicate that star-forming galaxies at these redshifts have high gas fractions \citep{tgn+10}, and dynamical masses are typically larger than stellar masses \citep{ess+06mass}.  A comparison of stellar and dynamical masses of $z\sim2$ galaxies with masses comparable to our LAE sample is presented by \citet{mvd+13}; these authors studied 14 objects with high equivalent width \OIII\ emission and found a median dynamical mass of $1.3\times10^9$ \msun\ and a stellar to dynamical mass ratio $M_{\star}/M_{\rm dyn} = 0.27$. If the LAEs in our sample are similar, we expect a median stellar mass of $M_{\star} = 1.8\times 10^8$ \msun.

\begin{deluxetable*}{l r r r r r r r r}
\tablewidth{0pt}
\tabletypesize{\footnotesize}
\tablecaption{{\it HST} Measurements and Dynamical Masses\label{tab:hst}}
\tablehead{
\colhead{} &
\colhead{} &
\multicolumn{2}{c}{ACS F625W} &
\multicolumn{2}{c}{ACS F814W} &
\multicolumn{2}{c}{WFC3 F160W} &
\colhead{} \\
\hline
\colhead{Object} &
\colhead{$z_{\rm neb}$\tablenotemark{a}} &
\colhead{$m_{\rm AB}$} &
\colhead{$a$\tablenotemark{b}} &
\colhead{$m_{\rm AB}$} &
\colhead{$a$\tablenotemark{b}} &
\colhead{$m_{\rm AB}$} &
\colhead{$a$\tablenotemark{b}} &
\colhead{$M_{\rm dyn}$\tablenotemark{c}} \\
\colhead{} &
\colhead{} &
\colhead{} &
\colhead{(kpc)} &
\colhead{} &
\colhead{(kpc)} &
\colhead{} &
\colhead{(kpc)} &
\colhead{($10^8$ \msun)}
}
\startdata
Q1700-BNB36 & 2.2949 & ... & ... & $25.54\pm0.02$ & $0.5\pm0.1$ & ... & ... &	$7.9\pm2.1$\\
Q1700-BNB47 & 2.2935 & ... & ... & $25.83\pm0.03$ & $0.6\pm0.1$ & ... & ... &	$3.5\pm1.8$\\
Q1700-BNB93 & 2.3254 & ... & ... & $26.89\pm0.09$ & $1.2\pm0.1$ & ... & ... &	$<11$\\
Q1700-BNB95 & 2.3064 & ... & ... & $25.69\pm0.03$ & $0.8\pm0.3$ & ... & ... &	$4.1\pm2$\\
Q1700-BNB104 & 2.2942 & ... & ... & $26.51\pm0.04$ & $<0.4$ & ... & ... &	$<1.6$\\
Q1700-BNB153 & 2.2903 & ... & ... & $26.63\pm0.12$ & $1.8\pm0.9$ & ... & ... &	$<13$\\
SSA22-001 & 3.0690 & $24.34\pm0.05$ & $1.1\pm0.2$ & $24.10\pm0.03$ & $1.2\pm0.1$ & ... & ... &	$68\pm6.7$\\
SSA22-004 & 3.0788 & ... & ... & $25.10\pm0.03$ & $1.3\pm0.5$ & ... & ... &	$40\pm15$\\
SSA22-006 & 3.0691 & ... & ... & $>29.6$ & ... & ... & ... & ...\\
SSA22-008 & 3.0692 & $25.10\pm0.14$ & $0.9\pm0.1$ & $24.75\pm0.04$ & $1.0\pm0.1$ & $24.44\pm0.05$ & $2.3\pm0.2$ &	$32\pm2.6$\\
SSA22-009 & 3.0687 & ... & ... & $25.26\pm0.08$ & $0.9\pm0.4$ & $25.11\pm0.07$ & $<0.9$ &	$6.4\pm3.2$\\
SSA22-013 & 3.0919 & ... & ... & $25.67\pm0.06$ & $0.9\pm0.5$ & ... & ... &	$14\pm11$\\
SSA22-014 & 3.0631 & $25.82\pm0.13$ & $<0.5$ & $25.89\pm0.04$ & $0.5\pm0.1$ & $25.40\pm0.09$ & $<0.9$ &	$2.5\pm1.5$\\
SSA22-021 & 3.0670 & ... & ... & $27.02\pm0.11$ & $<0.4$ & ... & ... &	$<6.3$\\
SSA22-046 & 3.0975 & $27.43\pm0.35$ & $<0.5$ & $27.20\pm0.11$ & $<0.4$ & $>27.54$ & ... &	$<1.3$\\
SSA22-062 & 3.0551 & $26.77\pm0.24$ & $<0.5$ & $26.77\pm0.08$ & $<0.4$ & ... & ... &	$<1.6$\\
SSA22-063 & 3.0978 & ... & ... & $27.09\pm0.12$ & $<0.4$ & $26.14\pm0.19$ & $<0.9$ &	$<2.9$\\
SSA22-066 & 3.0645 & $26.51\pm0.23$ & $<0.5$ & $26.87\pm0.08$ & $<0.4$ & $26.94\pm0.23$ & $<0.9$ &	$<6.1$\\
SSA22-067 & 3.1013 & ... & ... & $>29.6$ & ... & ... & ... & ...
\enddata
\tablenotetext{a}{Systemic redshift from \Ha\ or \OIII\ emission.}
\tablenotetext{b}{Major axis radius.}
\tablenotetext{c}{Calculated using size measured in F814W filter.}
\end{deluxetable*}

\subsection{Star formation rates}

We estimate star formation rates for the LAEs from their rest-frame UV continuum magnitudes traced by the $\cal{R}$-band filter, which corresponds to rest-frame wavelength 2100 \AA\ at $z=2.3$ and 1690 \AA\ at $z=3.1$. We use the prescription of \citet{k98}, converting from a \citet{s55} to a \citet{c03} initial mass function by dividing by a factor of 1.8. This conversion of UV luminosity to star formation rate is appropriate to the rest-frame UV luminosity between 1500 and 2800 \AA,  and applies only to populations with continuous star formation and ages of at least 100 Myr, since the UV luminosity is still rising for the first $\sim100$ Myr of an episode of star formation. We have no constraints on the ages of the LAEs, but given their low masses it would not be surprising if they were young objects. If this is indeed the case, the SFR calculated from the UV luminosity will be an underestimate of the true SFR.

We have not attempted to correct for dust extinction by measuring the UV slopes of these faint galaxies, so the SFRs calculated in this way should be considered lower limits for this reason as well. The extinction is likely to be low, however, since it is known to decrease with both decreasing luminosity \citep{bif+09} and increasing \lya\ emission \citep{ssp+03, vsat08,hos+09, kse+10,pgs+10,bag+11}. 

We find a median dust-uncorrected SFR of the LAE sample of 1.8 \msunyr. The minimum is 0.7 \msunyr, the maximum is 21 \msunyr, and 4 of the LAEs have SFR~$>10$ \msunyr.  Five of the LAEs in the Q1700 field also have \Ha\ fluxes measured with MOSFIRE. We compute the \Ha-based SFRs for these objects, again following the \citet{k98} prescription converted to a \citet{c03} initial mass function and without an extinction correction, and find that the \Ha\ and UV-based SFRs agree within a factor of 1.5 for 4 of the 5 objects, while the SFRs of the fifth differ by a factor of 1.7. These results are similar to those of \citet{ess+06stars}, who found that \Ha- and UV-based SFRs for individual galaxies agree to within a factor of $\sim2$. Given the significant uncertainties involved in the determination of star formation rates, we do not use the SFRs in the remainder of the paper. 

\subsection{Line ratios}

For the 5 LAEs with \Ha\ flux measurements, we can estimate the \lya\ escape fraction via the \lya/\Ha\ flux ratio.  The Q1700 galaxies BNB19, BNB36, BNB95, BNB104 and BNB153 have \lya/\Ha\ flux ratios of 3.2, 2.6, 3.1, 1.9 and 3.8, respectively; these values are factors of 2.2--4.4 lower than the typical value of 8.3 expected from Case B recombination \citep{fo85}. All 5 objects have thus lost a significant fraction of their \lya\ photons, either to differential slit losses due to spatial scattering of \lya\ \citep{sbs+11} or to destruction of \lya\ photons by dust. Within this small sample, there is no correlation between the \lya/\Ha\ ratio and the ${\cal R}$-band magnitude, or between the \lya/\Ha\ ratio and the ${\cal R}-{\mbox F814W}$ color for the four objects for which this color can be measured or limited. 

Because \OIII\ $\lambda 5007$ is the most commonly detected nebular emission line for the LAE sample, we can measure the ratio of the \lya\ and \OIII\ fluxes for 34 of the 36 LAEs. Values of \lya/\OIII\ vary widely, ranging from 0.1 (SSA22-006) to 5.2 (Q1700-BNB29); this wide variation is unsurprising, since the value of \lya/\OIII\ depends on the metallicity and ionization state of the gas in the \HII\ regions as well as on all of the factors influencing the escape of \lya.  \citet{gfg+13} have suggested that it may be possible to calibrate this ratio to primarily trace metallicity, ionization or \lya\ radiative transfer. We have little information on the metallicity or ionization state of the LAE sample, but we have tested the correlation of \lya/\OIII\ with the other quantities we have measured. The \lya/\OIII\ ratio is uncorrelated with both the ${\cal R}$-band magnitude and the offset of \lya\ emission from systemic velocity (discussed in Section \ref{sec:dv} below), but is correlated with the \OIII\ velocity dispersion with 3.1$\sigma$ significance, in the sense that galaxies with larger velocity dispersions have lower \lya/\OIII\ ratios.  However, this trend is driven entirely by an increase in \OIII\ flux with increasing velocity dispersion: the flux and velocity dispersion of \OIII\ are correlated with 3.6$\sigma$ significance, while the \lya\ flux is weakly correlated (2.0$\sigma$) with the \OIII\ flux and uncorrelated (0.27$\sigma$) with the velocity dispersion. We conclude that the \lya/\OIII\ ratio does not provide information unavailable from measurements of \OIII\ alone.  It may be worth revisiting this issue when more detailed measurements of ionization and metallicity are available, however, since both the opacity of the galaxy to \lya\ photons and the strength of \OIII\ emission are likely to be sensitive to the ionizing spectrum.

\begin{figure*}[htbp]
\plotone{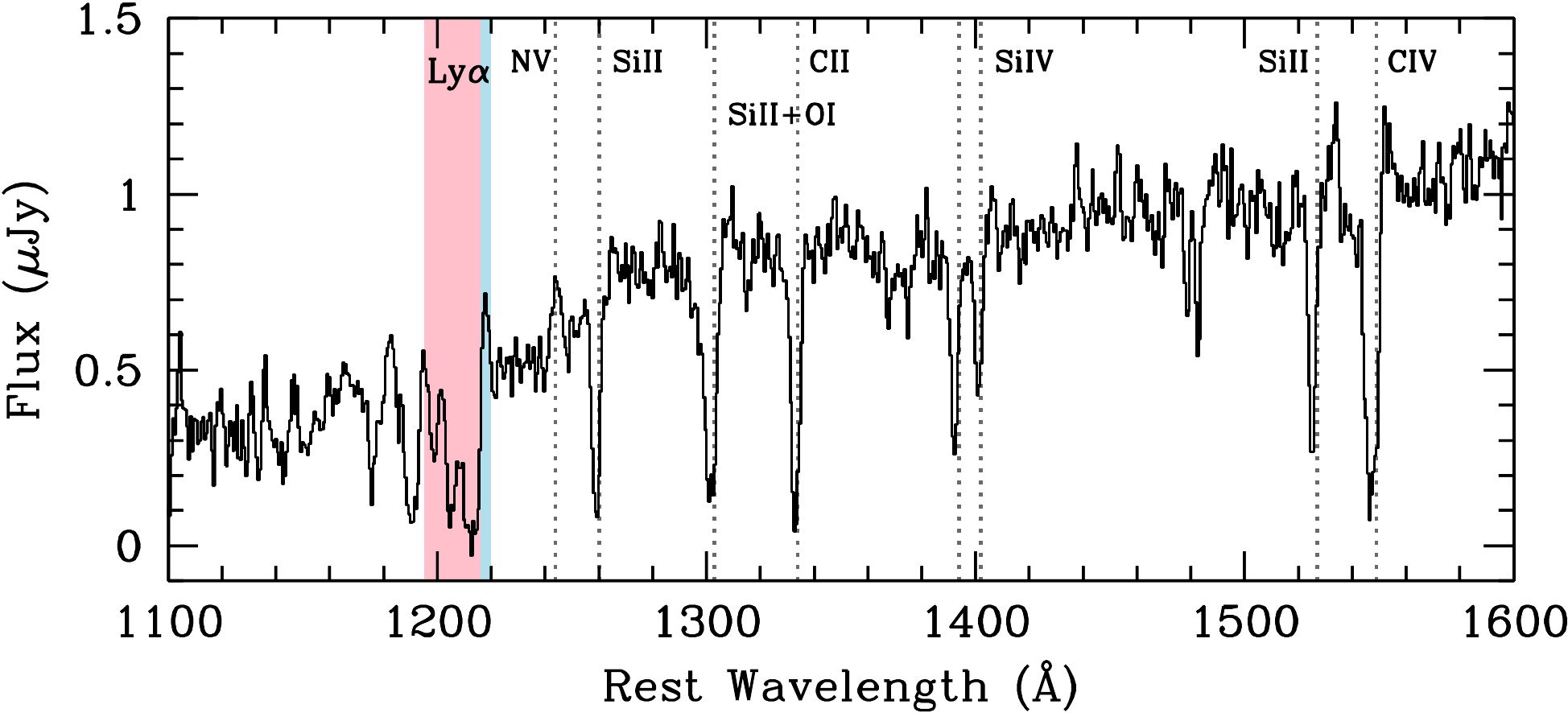}
\caption{An example of a galaxy in the comparison sample in which weak \lya\ emission (shaded in blue) is superimposed over strong absorption (shaded in red). This galaxy, Q2343-BX587 at $z=2.2429$, is typical in that a small emission peak is seen on the red side of the absorption line. As is usually the case,  the emission peak is highly redshifted; for this galaxy, \dvlya~$=564$ \kms\ while the total rest frame \lya\ equivalent width is \wlya~$=-8$ \AA, indicating net absorption. BX587 is also typical of such objects in having strong interstellar absorption lines (marked with dotted lines and labeled) and a red UV slope (see also \citealt{ssp+03}). We also mark and label \ion{N}{5} emission.}
\label{fig:bx587}
\end{figure*}

\section{The \lya\ Properties of Faint Galaxies at $z\sim2$--3}
\label{sec:profiles}

The \lya\ profiles of the galaxies in the LAE sample are shown in Figure \ref{fig:lyaprofiles}. They show a variety of spectral morphologies, ranging from a nearly symmetric single peak near zero velocity (e.g.\ Q1700-BNB42, SSA22-003) to purely redshifted emission (e.g.\ Q1700-BNB17, SSA22-006) to more complex, multi-component profiles (Q1700-BNB19, Q1700-BNB95).\footnote{The varying profiles are not an effect of varying spectral resolution, since as described in Section \ref{sec:opt}, the effective resolution at \lya\ is $\sim510$--530 \kms\ for all objects except the two (SSA22-014 and SSA22-046) observed at higher resolution.}  In this section we quantify the \lya\ velocity offset from the systemic velocity, the \lya\ equivalent width, and the fraction of emission emerging on the blue side of the systemic velocity for the LAEs and the comparison sample.  We report our measurements and the correlations between measured quantities here, and discuss the implications of our findings in Section \ref{sec:disc}.

\subsection{Ly$\alpha$\ velocity offsets}
\label{sec:dv}

With the benefit of precise systemic redshifts for the 36 LAEs and 122 comparison galaxies, we calculate the velocity offset of \lya\ emission with respect to the systemic velocity for the entire sample. We measure the flux-weighted centroid of each emission line to determine the \lya\ redshift, and calculate the velocity offset 
\begin{equation}
\Delta v_{\rm Ly\alpha} = c \left( \frac{z_{\lya}-z_{\rm sys}}{1+z_{\rm sys}} \right)
\label{eq:dvlya}
\end{equation}
where $z_{\rm sys}$ is the systemic redshift measured from the MOSFIRE spectra. For objects in which the \lya\ profile is pure emission (35 of the 36 galaxies in the LAE sample, as discussed in more detail in Section \ref{sec:wlya} below), we measure the continuum level on the blue and red sides of the line, and measure the centroid over the region in which the flux is higher than the continuum. The comparison sample also includes galaxies in which a weak \lya\ emission line is superimposed over strong absorption; an example of such an object is shown in Figure \ref{fig:bx587}, where the emission portion of the \lya\ profile is highlighted in blue and the absorption in red. In such cases we measure the velocity centroid of the emission portion of the line only, with the continuum set to its value redward of the emission line. In most cases the uncertainty in \dvlya\ is dominated by the uncertainty in the centroid of \lya, as determined from measurements of fake spectra described in more detail in Section \ref{sec:wlya}. For the LAE sample, the average uncertainty in the redshift of \lya\ is 65 \kms, while that of the systemic redshift from  MOSFIRE is 7 \kms.  Five objects in the sample have \dvlya\ uncertainty $>100$ \kms; these are galaxies with broad, multi-component emission lines (Q1700-BNB19 and Q1700-BNB95) or relatively weak features near residuals from sky line subtraction (SSA22-066, SSA22-072 and SSA22-078).  \lya\ redshifts and velocity offsets with uncertainties are reported in Table \ref{tab:lya}.

\begin{figure*}[htbp]
\plotone{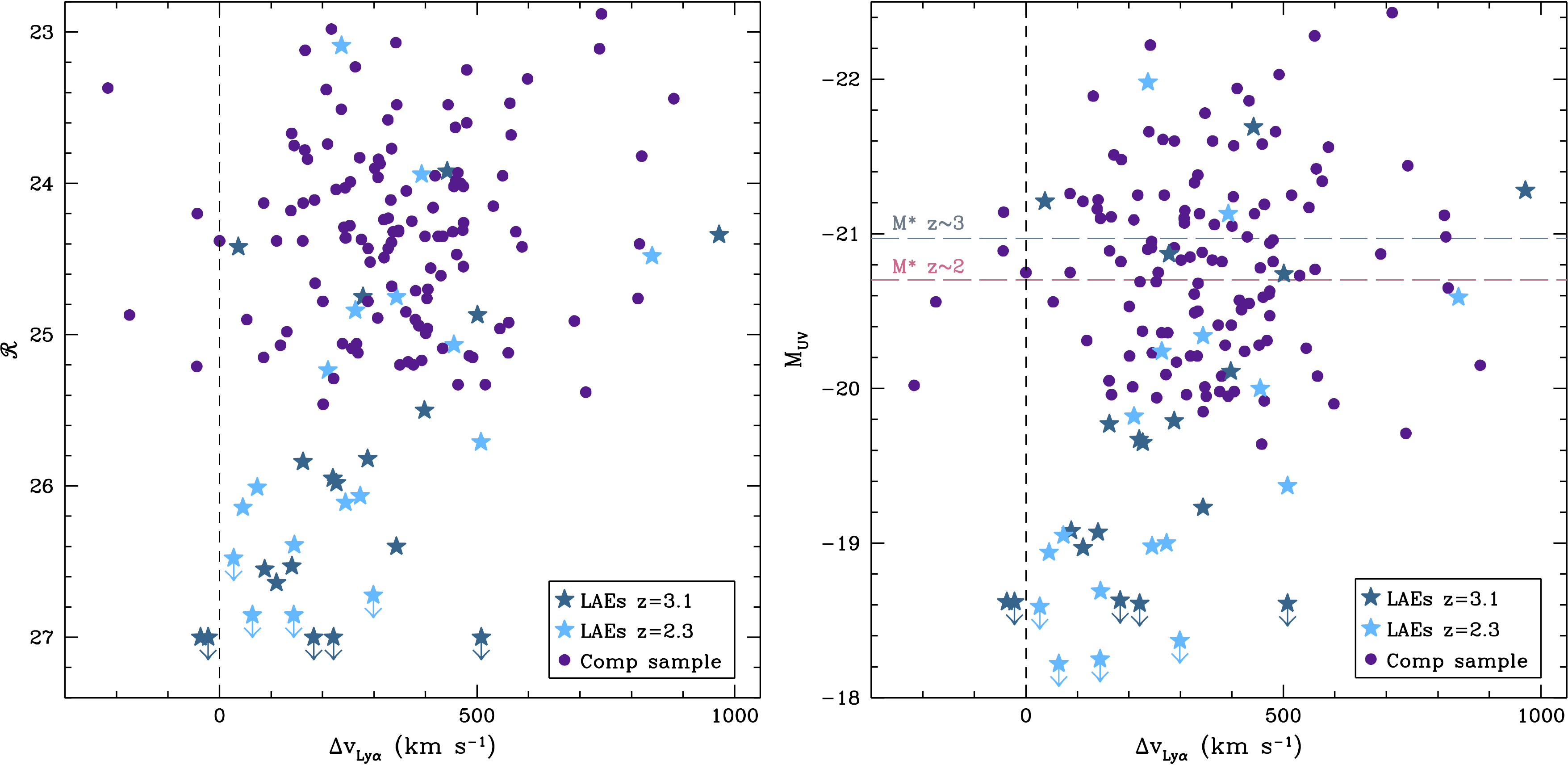}
\caption{{\it Left:} ${\cal R}$ magnitude vs.\ the velocity offset of \lya\ emission with respect to the systemic velocity measured from nebular emission lines. The LAEs are indicated by blue stars (dark blue for the $z\approx3.1$ SSA22 sample, and light blue for the $z\approx2.3$ Q1700 sample), and the comparison sample is shown with purple circles. {\it Right:} Absolute rest-frame UV magnitude $M_{\rm UV}$ vs.\ the velocity offset of \lya\ emission. The red and blue dashed horizontal lines indicate $M^*$ at $z\sim2$ and $z\sim3$ respectively \citep{rs09}. For galaxies in the LAE sample alone, \dvlya\ is anti-correlated with $\cal R$ and $M_{\rm UV}$ with 3.3$\sigma$ and 3.2$\sigma$ significance respectively (Spearman correlation coefficients $r_s=-0.52$ and $r_s=-0.51$).}
\label{fig:dvr}
\end{figure*}

In Figure \ref{fig:dvr} we plot the velocity offset of \lya\ emission with respect to systemic velocity against the continuum $\cal R$-band magnitude in the left panel and the rest-frame UV absolute magnitude $M_{\rm UV}$ in the right panel, where we have assumed a flat UV slope between $\sim1700$ and $\sim2100$ \AA. The 36 LAEs are shown as blue stars (the $z=3.1$ sample in dark blue, and the $z=2.3$ sample in light blue), and the 122 galaxies in the continuum-selected comparison sample are plotted with purple circles. We first note that the vast majority of the sample has \dvlya~$>0$ \kms, indicating the presence of outflowing gas in even the faintest galaxies in the sample. The \lya\ velocities span a wide range, however, from $<0$ \kms\ to nearly 1000 \kms. Both the LAEs and the comparison galaxies cover nearly this full range, but the LAEs tend to have smaller velocity offsets: the median $\Delta v_{\lya}$ of the full LAE sample is 237 \kms, while that of the comparison sample is 344 \kms. However, as we discuss below, this difference may be accounted for by the different continuum magnitude ranges of the two samples.

The LAEs range in magnitude from ${\cal R} \simeq 23$ to ${\cal R} > 27$. Considering this sample alone, it is clear from both panels of Figure \ref{fig:dvr} that fainter galaxies tend to have smaller \lya\ velocity offsets: dividing the sample in half at the median magnitude ${\cal R}=26$, we find that the median $\Delta v_{\lya}$ of the bright subsample is 344 \kms\ (identical to the median \dvlya\ of the comparison sample), while that of the faint subsample is 144 \kms.  A Spearman correlation test finds that $\Delta v_{\lya}$ and $\cal R$ are anti-correlated with 3.3$\sigma$ significance. We discuss the relationship between ${\cal R}$ and \dvlya\ for ease of comparison with other samples at similar redshifts, but because the LAE sample is composed of two subsamples at $z\sim2.3$ and $z\sim3.1$, a comparison of \dvlya\ and the rest-frame absolute magnitude $M_{\rm UV}$ is more physically meaningful. Dividing the LAEs at the median $M_{\rm UV}=-19.3$, we find median velocity offsets of 316 \kms\ for the bright subsample and 140 \kms\ for the faint subsample, and the Spearman test finds that $\Delta v_{\lya}$ and $M_{\rm UV}$ are anti-correlated with 3.2$\sigma$ significance. 

These results suggest that when LAEs are restricted to the brighter magnitudes typical of continuum-selected samples, their velocity offsets and those of continuum-selected galaxies are similar.  Our comparison sample is restricted to galaxies with ${\cal R}<25.5$; applying a similar restriction to the LAE sample, we find that the median $\Delta v_{\lya}$ of the 12 LAEs with ${\cal R}<25.5$ is 393 \kms. A two-sample K-S test also indicates that the two velocity distributions are indistinguishable; we find $p=0.92$, indicating that the null hypothesis that the velocity offsets of continuum-selected galaxies and LAEs with ${\cal R}<25.5$ are drawn from the same distribution cannot be rejected. This result is in contrast to that of \citet{son+14}, who found that even bright LAEs have systematically smaller velocity offsets than LBGs. This difference may be due to differing \lya\ equivalent width ranges in the two LAE samples, as discussed in Section \ref{sec:disc}.

The faint (${\cal R}>27$) galaxy SSA22-006 is an outlier among the faint LAEs, with a large velocity offset \dvlya~$=508$ \kms.  If this object is removed from the sample, the significances of the anti-correlations between \dvlya\ and $\cal R$ and \dvlya\ and $M_{\rm UV}$ increase to 4$\sigma$ and 3.8$\sigma$ respectively. SSA22-006 has the largest nebular line velocity dispersion $\sigma$ in the LAE sample: its value of $\sigma = 173 \pm 9$ \kms\ is more than three times higher than the sample average of 55 \kms, and 54\% higher than the next highest in the sample (SSA22-001, with $\sigma=112$ \kms).  SSA22-006 also has the largest ratio of nebular line flux to \lya\ flux in the sample, suggesting that its optical faintness may be due to greater extinction rather than lower mass.

\begin{figure}[htbp]
\plotone{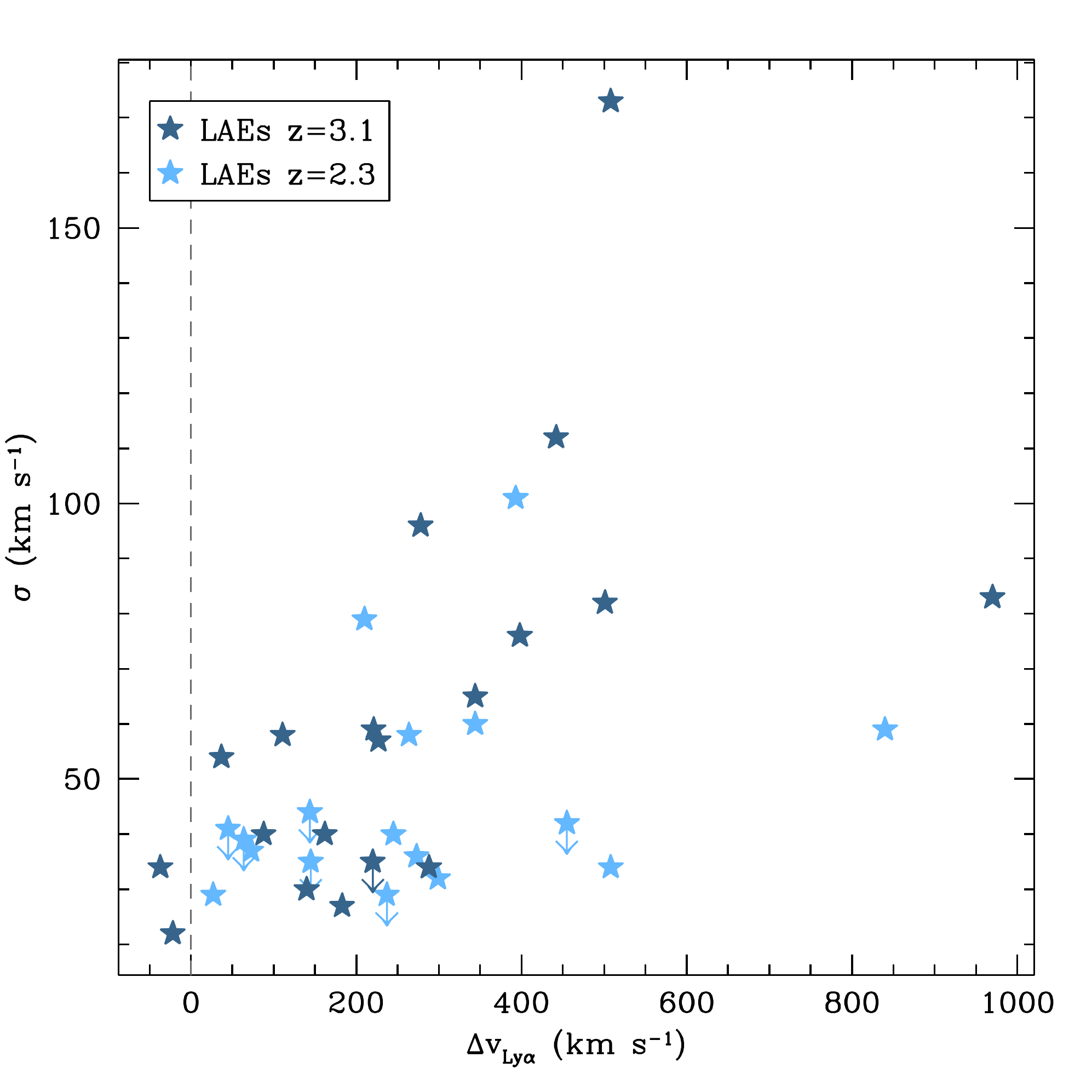}
\caption{The velocity dispersion $\sigma$ measured from nebular emission lines for the LAE sample vs.\ the velocity offset of \lya\ emission.  \dvlya\ and $\sigma$ are correlated with 3.4$\sigma$ significance, with Spearman correlation coefficient $r_s=0.54$.}
\label{fig:dvsig}
\end{figure}

Motivated in part by SSA22-006, we compare the velocity offset \dvlya\ with the velocity dispersion $\sigma$, measured from the width of \OIII$\lambda5007$ emission (except for Q1700-BNB104 and Q1700-BNB153, for which we use \Ha).  This comparison is shown  in Figure \ref{fig:dvsig}. Although there is considerable scatter, especially among galaxies with larger values of \dvlya\ and $\sigma$, we find that galaxies with larger velocity dispersions also tend to have larger \lya\ velocity offsets; a Spearman test finds that \dvlya\ and $\sigma$ are correlated with 3.4$\sigma$ significance.
 
We conclude that, among galaxies selected via \lya\ emission, the velocity offset of \lya\ increases with both increasing velocity dispersion and increasing brightness, suggesting a relationship with both galaxy mass and star formation rate. These relationships have already been suggested by \citet{son+14}, who find a correlation between \dvlya\ and the stellar mass from SED fitting in their smaller, brighter LAE sample; they also report a weak (1.8$\sigma$) correlation between \dvlya\ and SFR. Finally, we note that these correlations do not appear to be present in the brighter comparison sample, although this may be due to a lack of dynamic range in mass and luminosity as well as the more significant effects of dust in brighter galaxies. 
 
\begin{figure*}[htbp]
\plotone{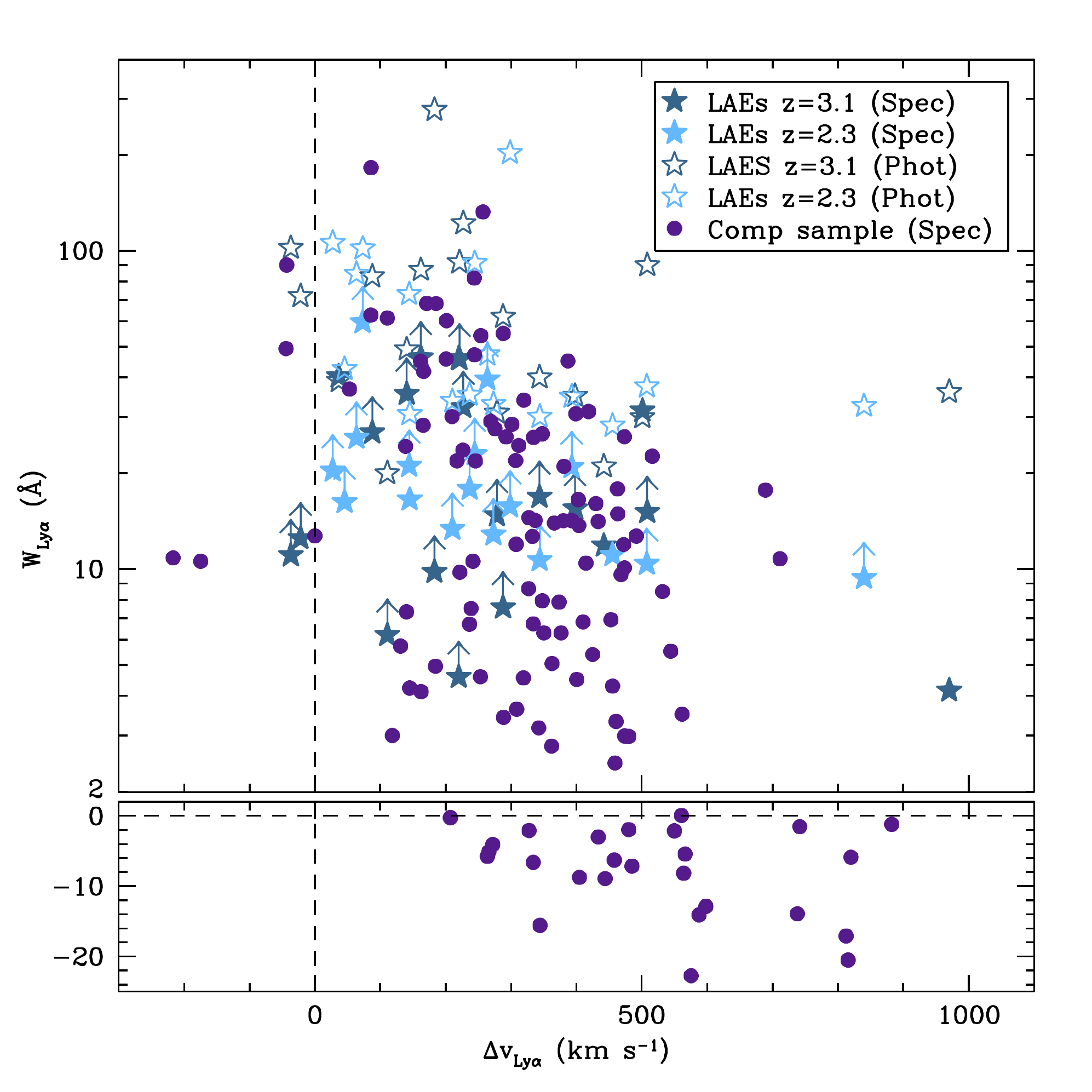}
\caption{\lya\ rest-frame equivalent width vs.\ the velocity offset of \lya\ emission with respect to the systemic velocity measured from nebular emission lines. The LAEs are indicated by blue stars (dark blue for the $z\approx3.1$ SSA22 sample, and light blue for the $z\approx2.3$ Q1700 sample), and the comparison sample is shown with purple circles. For the LAE sample, equivalent widths measured from narrowband imaging are shown with open stars, and equivalent widths measured from the \lya\ spectra are shown with filled stars. For the comparison sample we measure the spectroscopic equivalent width only. Among the total sample of 158 individual galaxies,  \wlya\ and \dvlya\ are anti-correlated with 6.7$\sigma$ significance ($r_s=-0.50$) when the spectroscopic equivalent widths (with limits treated as detections) are used for the LAE sample, and with 7.6$\sigma$ significance ($r_s=-0.56$) when the photometric equivalent widths are used.}
\label{fig:dv_wlya}
\end{figure*}

\subsection{Ly$\alpha$\ equivalent widths}
\label{sec:wlya}

We measure spectroscopic \lya\ equivalent widths for the LAE and comparison samples using a slight modification of the procedure described by \citet{kse+10}. The method is described in detail in Section 2.3 of \citet{kse+10}; we review it briefly here, and focus on the differences in our method. The most important of these differences is due to the fact that the continuum is generally not detected for the faint LAEs, meaning that most of our spectroscopic equivalent width measurements for this sample are lower limits. 

In brief, the \lya\ profile of each galaxy is classified as either pure emission or as a combination of emission and absorption (\citealt{kse+10} also employed pure absorption and noise classifications, but because all the galaxies in our sample are selected to have \lya\ in emission, these classifications are not present in the current work). One of the most extreme examples of a combination object is shown in Figure \ref{fig:bx587}. Among the LAEs, 35 are classified as emission and one (SSA22-004, which is also the Lyman break galaxy SSA22-MD23; see \citealt{nsss11}) as combination, while the comparison sample contains 82 emission and 40 combination objects.

We next measure the red and blue continuum levels on either side of the \lya\ emission line, measure the flux in the line, and calculate the equivalent width to be the line flux divided by the red continuum level; these steps are described in detail by \citet{kse+10}. We calculate uncertainties in the equivalent width measurements by perturbing each spectrum 100 times by an amount drawn from a Gaussian distribution with standard deviation equal to the RMS of the red continuum, measuring the equivalent width of each fake spectrum, and calculating the average and standard deviation of the equivalent width measurements. We adopt this average and standard deviation as the equivalent width measurement and its error, except in cases where we determine limits as described below. These perturbed spectra are also used to determine the uncertainties in \lya\ velocity offsets discussed in Section \ref{sec:dv}.

Our ability to measure equivalent widths for the LAEs is limited by the S/N of the continuum, which is usually very low. For most of the LAE sample, we adopt 1$\sigma$ lower limits on the equivalent width, calculated as follows. For each object, we calculate the equivalent width to be the average of the line flux divided by the continuum level for the 100 fake spectra, as described above. We also calculate the average of the line flux divided by the continuum uncertainty (given by the RMS of the continuum) for the fake spectra. If the equivalent width is higher than the flux divided by continuum RMS measurement, we consider the equivalent width measurement to be a detection; this is equivalent to requiring continuum S/N$~>1$, and applies to 6 of the 36 LAEs. If the equivalent width is lower than the flux divided by continuum RMS measurement, we adopt the flux/(continuum RMS) measurement as the 1$\sigma$ lower limit on the equivalent width.  In all cases these lower limits are smaller than the equivalent widths determined from narrow-band imaging; see Table \ref{tab:lya}.\footnote{Note that the spectroscopic equivalent width measurements are likely to be smaller than equivalent widths measured from narrow-band imaging even when both are well-detected; using composite spectra and imaging of a sample of 92 galaxies at $z\sim2.65$, \citet{sbs+11} find that the spectroscopic  equivalent width is on average $\sim5$ times lower than the photometric equivalent width, due to the spatial scattering of \lya\ photons into an extended halo. Further studies have suggested that the spatial extent of such halos may depend on galaxy type and environment \citep{myh+12,mon+14}.} As we will see below, few conclusions can be drawn from the equivalent widths of the LAE sample alone, given these limits; the more robust equivalent width measurements of the comparison sample are needed for meaningful results.

In Figure \ref{fig:dv_wlya} we plot the \lya\ equivalent width against the velocity offset \dvlya. There is an anti-correlation between \wlya\ and \dvlya\ such that galaxies with higher equivalent width \lya\ emission tend to have smaller velocity offsets. Such an anti-correlation has been previously reported by \citet{hos+13} and \citet{son+14} using samples of 10--20 LAEs and a binned average of 41 LBGs; we confirm the trend here with 158 individual galaxies, combining the LAEs and comparison sample in order to increase the dynamic range in \wlya. Using a Spearman test to compare \wlya\ and \dvlya\ for the combined sample of 122 comparison objects and 36 LAEs, and using the spectroscopic equivalent width measurements with limits treated as detections, we find that the probability that the two quantities are uncorrelated is $P=2\times10^{-11}$, giving a significance of $6.7\sigma$. If instead we use the photometric \wlya\ measurements from narrow-band imaging for the LAE sample, the probability of non-correlation decreases further to $P=3\times10^{-14}$ ($7.6\sigma$). This second test has the disadvantage that the equivalent widths are measured in different ways for the two samples, but the advantage that the LAE measurements are detections rather than limits. 

We also note that the LAEs are not required to detect this correlation: in the comparison sample alone, \dvlya\ and \wlya\ are anti-correlated with $6.8\sigma$ significance.  On the other hand, it is much more difficult to detect such a trend using the LAE sample alone. There is no  correlation (0.8$\sigma$) between the spectroscopic \wlya\ and \dvlya\ for the LAE sample alone; however, given that most of the spectroscopic equivalent width measurements for the sample are generous lower limits, this is not surprising. We find a somewhat stronger correlation, with 2.6$\sigma$ significance, when comparing velocity offsets with the photometric \lya\ equivalent widths  (the open stars in Figure \ref{fig:dv_wlya}). We note that these two quantities are determined entirely independently, from spectroscopy and imaging respectively.  This result suggests that such a correlation may also exist among LAEs alone, but it is clear that the larger dynamic range in \wlya\ gained by including continuum-selected galaxies allows for a much more robust characterization of the trend.

An additional interesting question is whether fainter galaxies tend to have higher \lya\ equivalent widths; this may follow from the results that fainter galaxies tend to have lower values of \dvlya, and low values of \dvlya\ are associated with high \lya\ equivalent width. We are unable to test this with our current data, because all of the galaxies in our sample with ${\cal R} > 25.5$ were selected via their strong \lya\ emission. However, results from other studies suggest that the fraction of galaxies with strong \lya\ emission does indeed increase in fainter samples \citep{aoi+06,seo11,sds11}.

\begin{figure}[htbp]
\plotone{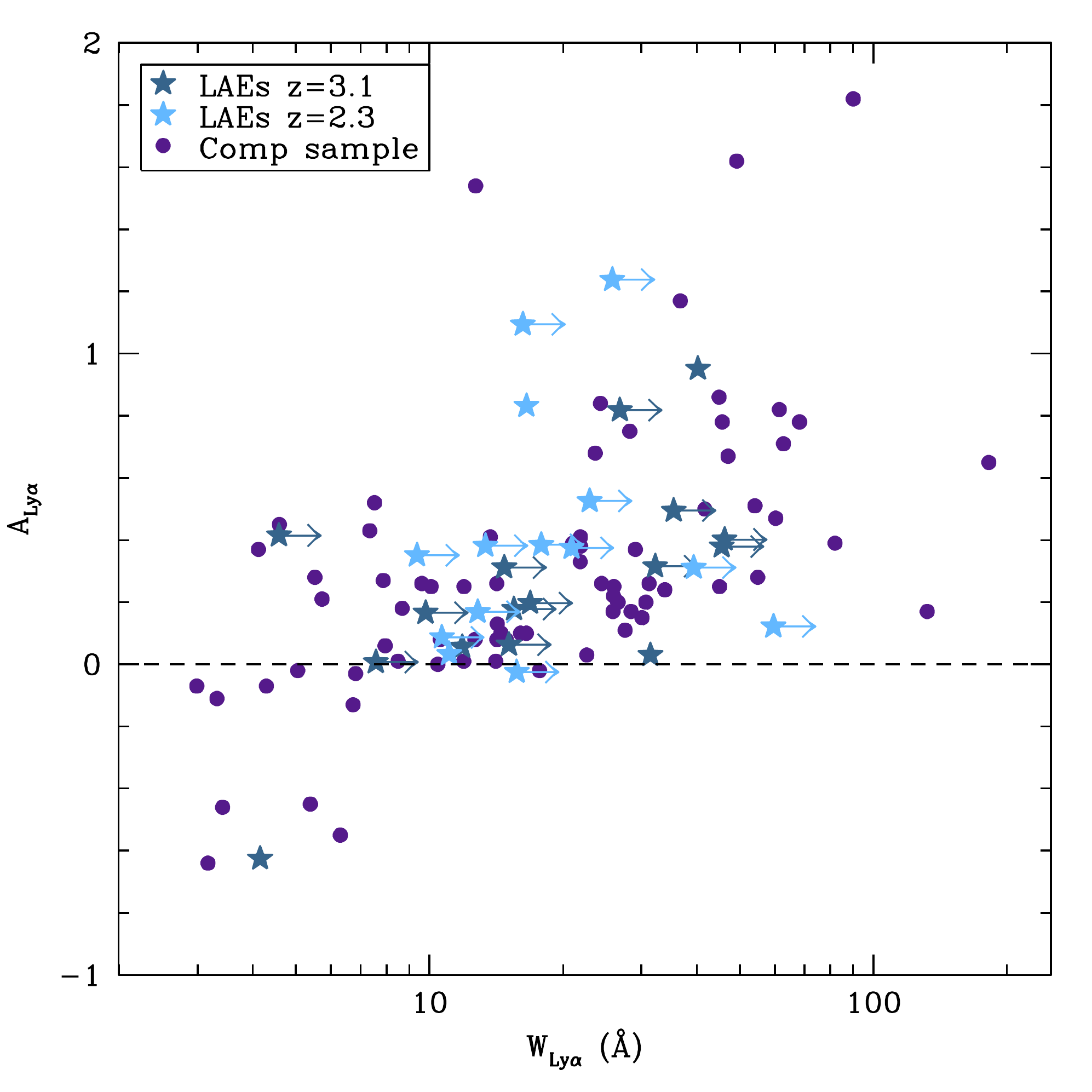}
\caption{The ratio of the \lya\ equivalent widths blueward and redward of zero velocity \Alya\ vs.\ \lya\ equivalent width. Only objects with uncertainties in \Alya\ less than 0.3 are shown. The LAE with \Alya~$=-0.6$ is SSA22-004, the only object in the LAE sample with \lya\ absorption blueward of the systemic velocity. Symbols are as in Figure \ref{fig:dvr}. The strength of the correlation between \Alya\ and \wlya\ is $6.3\sigma$ ($r_s=0.57$).}
\label{fig:asym_wlya}
\end{figure} 

\subsection{Blueshifted Ly$\alpha$\ emission}
\label{sec:asym}

Motivated by the detection of strong \lya\ emission with a significant component emerging blueward of the systemic velocity in both low mass, low metallicity galaxies at \ztwo\ (\citealt{eps+10}, Erb et al.\ in prep) and in \zthree\ galaxies with spectroscopic detection of escaping Lyman continuum photons (C. Steidel et al, in prep, but c.f.\ \citealt{nsss11,msn+13}), we estimate the fraction of \lya\ emission emerging at $v<0$ in the LAEs and comparison sample. Given the relatively low resolution of the \lya\ spectroscopy, we do not attempt a detailed modeling of the line profile.  We instead define the quantity $A_{\lya}$, which measures the asymmetry of the line with respect to the systemic redshift. We compare the equivalent widths of the line on either side of zero velocity: $A_{\lya} \equiv  W_{\lya\rm (blue)}/W_{\lya\rm (red)}$, where $W_{\lya\rm (blue)}$ and $W_{\lya\rm (red)}$ are measured by dividing the equivalent width measured in Section \ref{sec:wlya} into two portions at the rest wavelength of \lya. A line symmetric about zero velocity would then have $A_{\lya}=1$, while a purely redshifted line would have $A_{\lya}=0$; thus \Alya\ is highly correlated with the \lya\ velocity offset \dvlya. Negative values of $A_{\lya}$ are also possible, for objects with blueshifted absorption and redshifted emission (our sample does not contain objects with redshifted \lya\ absorption). Note that because the uncertainty in the continuum level cancels out, we are able to measure \Alya\ for the LAEs with only limits on the total equivalent width (although in practice, the uncertainties in \Alya\ are large for some of these objects).

Because most of the \lya\ emission lines are unresolved, the value of \Alya\ is clearly affected by the spectral resolution: a slightly redshifted emission line observed with increasingly lower spectral resolution will have an increasingly large blue fraction, as the line is spread over a wider range of velocities. This issue is further complicated by the fact that the resolution is not the same for all of the objects in the sample, since the Q1700 sample was observed with the 400-line grism while the SSA22 sample was observed with the 300-line grism (with the exception of two objects observed with the 600-line grism). However, the difference in resolution due to instrumental setup between the two samples is largely compensated for by the different observed wavelength of \lya\ at $z=2.3$ and $z=3.1$. The effective FWHM for the 300-line grism at $\sim 4985$ \AA\ is $\sim 530$ \kms, while that of the 400-line grism at $\sim4010$ \AA\ is $\sim 510$ \kms. This $\sim4$\% difference in resolution is small compared to the uncertainties in \Alya\ due to limited signal-to-noise, and we conclude that while the \textit{absolute} values of \Alya\ are certainly influenced by the limited spectral resolution, the \textit{relative} values are not. In other words, higher spectral resolution is required to obtain an accurate measurement of the fraction of emission emerging blueward of zero velocity, but objects with higher values of \Alya\ in our sample have a higher fraction of blue emission than those with lower values.

We plot \Alya\ against the spectroscopic \lya\ equivalent width \wlya\ in  Figure \ref{fig:asym_wlya}, including only objects with uncertainties in \Alya\ less than 0.3.  We note here that there are two objects (SSA22-072 and SSA22-078) in the LAE sample with strong, blueshifted peaks at $-800$ to $-1000$ \kms\ in their \lya\ spectra (see Figure \ref{fig:lyaprofiles}). Neither of these peaks contribute to our measurements of \Alya:\ SSA22-072 is not included in the sample because the uncertainty in \Alya\ is too large, and the blue peak in the spectrum of SSA22-078 is not measured as part of the \lya\ profile, since the spectrum reaches the continuum level between the two peaks (unlike the other multiple-peaked profiles in the sample).

We observe a correlation between \Alya\ and \wlya, such that objects with stronger \lya\ emission have a larger fraction of flux emerging blueward of systemic velocity. Applying a Spearman test to the combined sample of LAEs and comparison objects, we find that the probability of no correlation is $P=2\times10^{-10}$ ($6.3\sigma$), while a test of the comparison sample alone finds $P=4\times10^{-9}$ ($5.9\sigma$). Among the LAE sample alone, we find no significant correlation between \Alya\ and \wlya, when either the spectroscopic or photometric equivalent widths are used.  As with \wlya\ and \dvlya\ (Section \ref{sec:wlya}), the LAE sample is not required in order to detect correlation between \lya\ properties. 

\begin{figure}[htbp]
\plotone{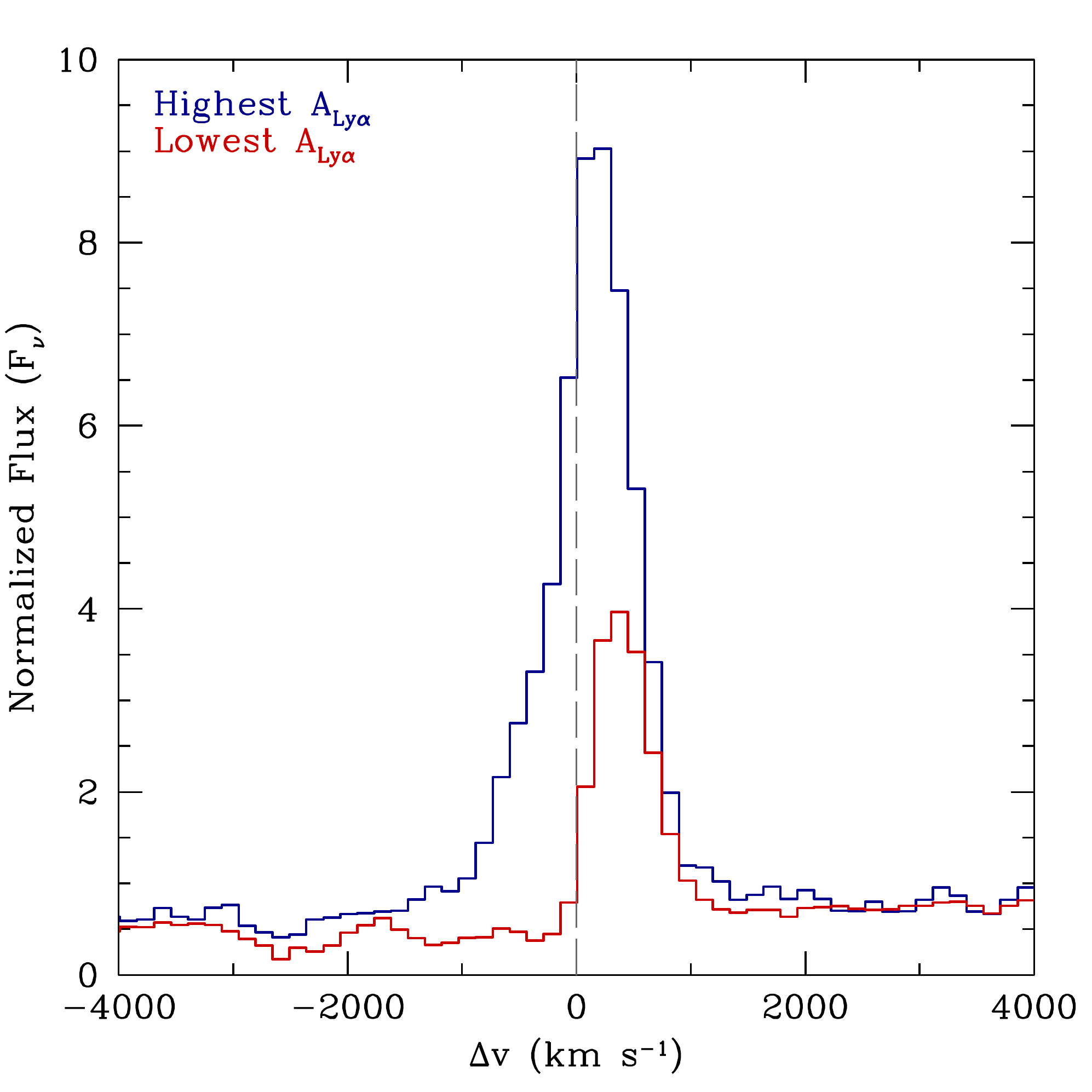}
\caption{Composite \lya\ profiles of the upper and lower thirds of the comparison sample, as measured by the \lya\ asymmetry \Alya. Galaxies with the highest values of \Alya\ have the most emission emerging blueward of zero velocity and are shown in blue; the red line shows the composite spectrum of galaxies with the lowest values of \Alya\ and the lowest fraction of blueshifted emission.}
\label{fig:redblue_lya}
\end{figure}

In order to more directly assess the relationship between \wlya, \Alya\ and outflow velocity we turn to the comparison sample. We have no direct information about outflow velocities among the LAEs, since absorption lines are not detected in the vast majority of the sample or in the composite spectrum of the 36 LAEs (but see R.~Trainor et al.\ in prep). We can, however, detect absorption lines in most of the brighter comparison galaxies. We divide the comparison sample into three bins based on \Alya, again using only objects with uncertainties in \Alya\ less than 0.3. The 25 galaxies with the highest values of \Alya\ show significant blueshifted emission, while the \lya\ emission of the 25 galaxies with the lowest values of \Alya\ is purely redshifted. We construct composite spectra of the galaxies in the high and low \Alya\ bins by averaging the spectra, scaling each individual spectrum to its median value between 1250--1300 \AA\ and rejecting the 3 highest and lowest pixels. We then normalize the average spectrum by its continuum value redward of \lya. The \lya\ profiles of these composite spectra are shown in Figure \ref{fig:redblue_lya}. As also shown by the individual galaxies in Figure \ref{fig:asym_wlya}, galaxies with significant blue emission have stronger \lya\ emission at all wavelengths. The \lya\ equivalent width of the highest \Alya\ subsample is 41 \AA, while that of the lowest \Alya\ subsample is 16 \AA.

\begin{figure}[htbp]
\plotone{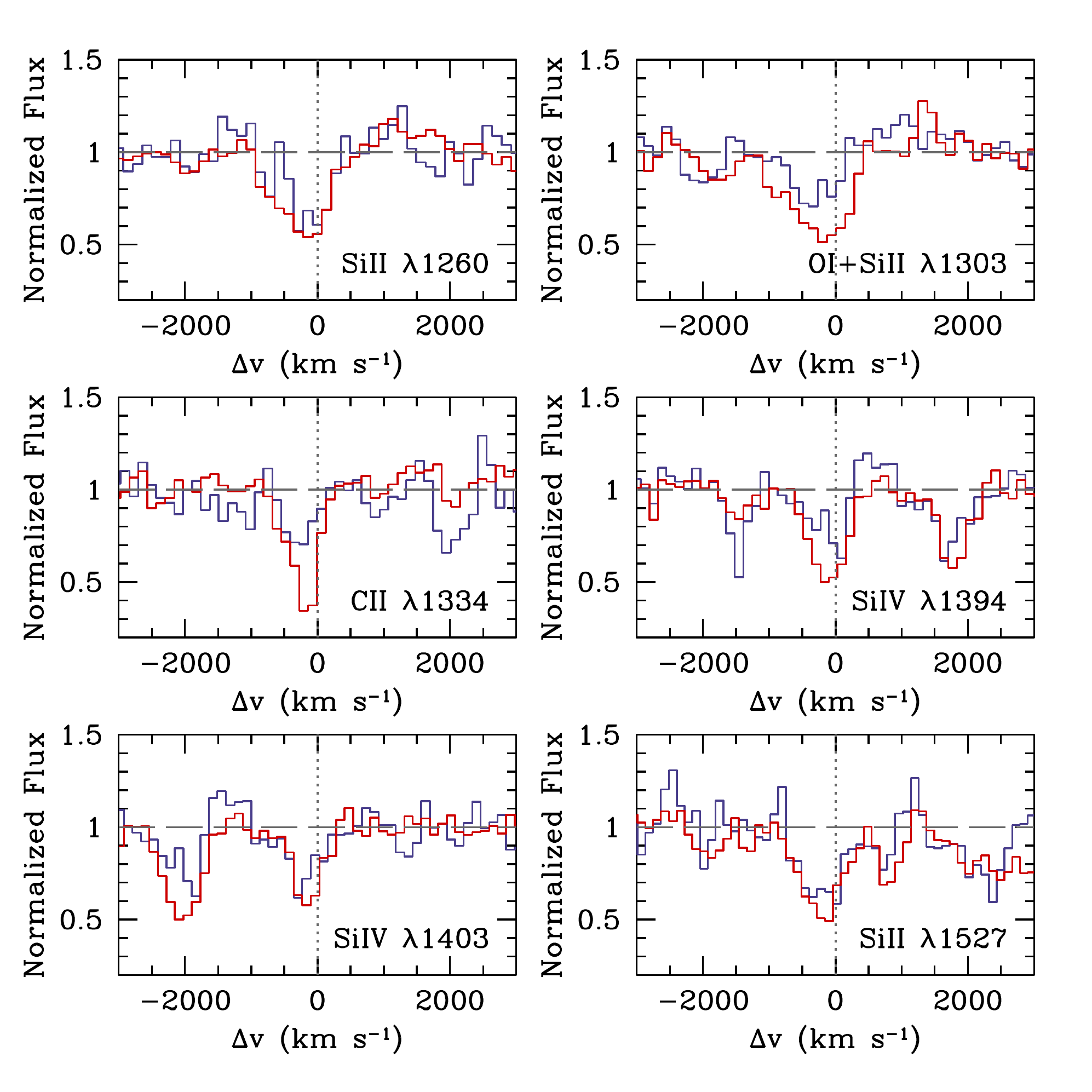}
\caption{Composite absorption line profiles of the upper and lower thirds of the comparison sample, as measured by the \lya\ asymmetry \Alya. Galaxies shown in blue have the highest values of \Alya\ and the most emission emerging blueward of zero velocity, and the red line shows the composite spectrum of galaxies with the lowest values of \Alya\ and the lowest fraction of blueshifted emission.  Objects with lower values of \Alya\ have stronger absorption lines, but no significant absorption line velocity differences.}
\label{fig:redblue_abslines}
\end{figure}

A closer look at the two profiles shows that their red wings are very similar; the average velocity offset \dvlya\ given by the line centroid 
is obviously smaller for the profile with more blueshifted emission, but this is caused by the increased emission on the blue side of the line rather than by an overall shift to bluer wavelengths.  This difference suggests that galaxies with more symmetric profiles and thus higher \Alya\ and stronger \wlya\ have a lower optical depth to \lya\ photons near the systemic redshift, rather than lower outflow velocities. We test this theory by examining the absorption lines of the high and low \Alya\ composites, as shown in Figure \ref{fig:redblue_abslines}. Here we plot the low ionization transitions \ion{Si}{2} $\lambda$1260, \ion{O}{1}~+~\ion{Si}{2} $\lambda$1303,  \ion{C}{2} $\lambda$1334 and  \ion{Si}{2} $\lambda$1527, and the high ionization transitions  \ion{Si}{4} $\lambda\lambda$1394, 1403. We find that the spectrum of galaxies with low \Alya\ and purely redshifted \lya\ emission (plotted in red) has stronger absorption in both the low and high ionization transitions than the high \Alya\ spectrum (plotted in blue). These results are not unexpected, given the previously known anti-correlation between \lya\ equivalent width and the strength of interstellar absorption lines \citep{ssp+03}.  Absorption lines in low resolution spectra represent a somewhat crude probe of outflow velocities, since they are a superposition of absorption from outflowing gas and gas at the systemic redshift (e.g.\ \citealt{wcp+09,ses+10,msc+12}).  However, there are no significant differences in either the centroids or the blue wings of the absorption lines between the high and low \Alya\ spectra, indicating that (at least for the bright galaxies in the comparison sample) the differences in the \lya\ profile are likely to be due to the optical depth of neutral gas rather than to differences in outflow velocity.

\section{Summary and Discussion}
\label{sec:disc}
We have studied the \lya\ properties of 36 LAEs at \ztwo--3 with magnitudes ${\cal R} \sim 23$ to ${\cal R} > 27$ and a comparison sample of 122 UV-color-selected galaxies with ${\cal R} < 25.5$, all with \lya\ emission. All 158 galaxies have systemic redshifts measured from nebular emission lines.  Seventeen of the 36 LAEs fall within {\it HST} imaging fields and are detected with F814W magnitudes ranging from 24.1 to 27.2.  We use the sizes measured in the F814W filter in combination with nebular line widths to estimate dynamical masses, finding a median dynamical mass of $6.3\times10^8$ \msun, subject to a factor of $\sim2$ systematic uncertainty due to the unknown mass distribution.  We consider the \lya\ velocity offsets from systemic \dvlya, the \lya\ equivalent widths \wlya, and the ratios of the \lya\ equivalent widths blueward and redward of zero velocity \Alya, as well as the $\cal R$-band magnitudes, rest-frame UV absolute magnitudes $M_{\rm UV}$, and nebular velocity dispersions $\sigma$. Our primary results are as follows:

\begin{itemize}
\item{The velocity offset \dvlya\ of the LAE sample is anti-correlated with $\cal R$-band apparent magnitude and rest-frame UV absolute magnitude $M_{\rm UV}$ with $>3\sigma$ significance:\ continuum-bright LAEs tend to have larger velocity offsets than continuum-faint LAEs, and the \dvlya\ distribution of LAEs with ${\cal R} < 25.5$ is indistinguishable from that of continuum-selected galaxies in the same magnitude range (see Figure \ref{fig:dvr}).}  
\item{Among the LAE sample, \dvlya\ and nebular line velocity dispersion $\sigma$ are correlated with $>3\sigma$ significance:\ galaxies with higher velocity dispersions tend to have larger \lya\ velocity offsets relative to the systemic velocity (see Figure \ref{fig:dvsig}).} 
\item{Using the full sample of 158 galaxies, we find that \dvlya\ is anti-correlated with the \lya\ equivalent width with $7\sigma$ significance:\ galaxies with higher equivalent widths tend to have smaller velocity offsets. The large dynamic range in \wlya\ provided by the comparison sample is required to detect this trend; the correlation is nearly as strong in the comparison sample alone, while correlations between \dvlya\ and \wlya\ in the LAE sample alone are marginal (see Figure \ref{fig:dv_wlya}).} 
\item{Galaxies with higher equivalent width \lya\ emission also have  larger values of \Alya, indicating a higher fraction of \lya\ photons emitted blueward of the systemic velocity (see Figure \ref{fig:asym_wlya}).}  
\item{Galaxies with larger values of \Alya\ naturally have smaller \lya\ velocity offsets. For bright, continuum-selected galaxies for which absorption lines can be measured, we find no evidence that this shift in \dvlya\ is related to the outflow velocity; galaxies with higher \Alya\ have weaker interstellar absorption lines, but no significant velocity difference in either the centroid or the blue wing of the absorption lines (see Figures \ref{fig:redblue_lya} and \ref{fig:redblue_abslines}).}  
\end{itemize}
We discuss the implications of these results below.

\subsection{The \lya\ velocity offset and the velocity of galactic outflows}

We have seen that galaxies in the LAE sample have \lya\ velocity offsets \dvlya\ that decrease with both decreasing luminosity and decreasing velocity dispersion. An important question is whether or not these trends are indicative of lower outflow velocities in fainter, lower mass LAEs. Such a result would not be surprising, since lower outflow speeds (as measured from the centroids of interstellar absorption lines) are observed in galaxies with lower masses and lower star formation rates and SFR densities, both locally \citep{m05,rvs05,hbo+11} and at intermediate and higher redshifts \citep{wcp+09,dmt+12,ksm+12}.  The theoretical expectation is also that more massive galaxies with higher star formation rates will be able to drive faster outflows. Recent simulations of galactic outflows find that, while outflowing gas covers a wide range of velocities in galaxies of all types, the typical velocity beyond which the amount of outflowing material drops off rapidly is higher in massive gas-rich starbursts than in dwarf galaxies. The relationship between this outflow turnover velocity and galaxy circular velocity is not monotonic, however, and the differences in outflow velocity are probably due to the relative importance of radiation pressure and thermal heating in driving the outflow \citep{hqm12}.

However, there is substantial evidence that factors other than the speed of outflowing gas are crucial in setting the velocity offset \dvlya.  Radiative transfer modeling indicates that the column density and covering fraction of neutral gas, the inclination of the galaxy, and the dust content all strongly influence the emergent \lya\ profile (e.g.\ \citealt{vsm06,vsat08,vdb+12,ksk+12}). More specifically, increasing the column density of neutral hydrogen in a simple expanding shell model shifts the peak of \lya\ emission farther to the red, even if the outflow velocity is unchanged \citep{vsm06}. \citet{ses+10} offer an alternative model for \lya\ emission, in which the line profile is largely determined by the bulk velocity and covering fraction of clumps of outflowing gas. In this case the covering fraction is more important than the column density, since the \lya\ photons generally do not penetrate the dense clouds of gas. This model also shows that gas near the systemic velocity has a strong effect on the line profile:\ increasing the amount of gas near zero velocity shifts the peak of \lya\ emission to the red, as the photons then require larger velocity shifts in order to escape the galaxy.

More recently, \citet{cbh+13} have modeled the high-resolution spectra of three bright LAEs, emphasizing that the column density of neutral hydrogen is the dominant factor in determining the \lya\ profile, and that therefore the \lya\ velocity offset should not be equated with the outflow velocity, particularly when the offset is measured from low resolution spectroscopy. Similarly, \citet{son+14} find that $N_{\rm HI}$ is a primary factor in determining the \lya\ emissivity.  While spectra of higher resolution and higher S/N are required in order to reliably distinguish the effects of variations in covering fraction and $N_{\rm HI}$, these results are in general agreement with our findings in Section \ref{sec:asym}, in which we show that, at least for bright galaxies, smaller values of \dvlya\ can arise from a larger fraction of emission emerging blueward of systemic velocity, while the red wing of the \lya\ profile and the outflow velocity  traced by absorption lines remain unchanged. Our results can be effectively summarized by the main conclusions of  \citet{ses+10}:\ the peak velocity of \lya\ emission is primarily influenced by the properties of gas at the systemic velocity of the galaxy, while the red wing traces outflowing gas at the largest velocities for which the covering fraction is high enough to scatter an appreciable number of photons.

We therefore conclude that, while it is likely that the faint LAEs in our sample do tend to have lower outflow speeds, there is little evidence for this from the \lya\ velocity offsets, since the \lya\ profile is likely to be strongly affected by other physical conditions within the galaxy. More direct measurements of outflow velocity from absorption lines are required to address this question.

\subsection{\lya\ velocity offset and equivalent width}

In Section \ref{sec:wlya} we showed that galaxies with stronger \lya\ emission tend to have smaller \lya\ velocity offsets. Although we have presented the strongest measurement of this correlation to date, with measurements of 158 individual galaxies over a wide range in mass and with 7$\sigma$ significance, the idea that a decrease in the equivalent width of \lya\ emission is accompanied by a shift to higher velocities is not new. \citet{mkt+03} showed that increasing the column density of an expanding shell of gas both decreases the strength of \lya\ emission and shifts the peak to redder wavelengths.  The models of \citet{ses+10} also show that increasing the amount of gas near zero velocity can both decrease the \lya\ equivalent width and shift the peak to the red. Earlier observations have also suggested this trend. Using composite spectra of Lyman break galaxies binned according to \lya\ equivalent width, \citet{ssp+03} found that the velocity offset between \lya\ emission and the interstellar absorption lines decreases with increasing \lya\ equivalent width, and indirect determinations of systemic redshifts suggested that this difference was due to a decrease in \dvlya.  More recently, \citet{hos+13} and \citet{son+14} have found an anti-correlation between \dvlya\ and \wlya\ using samples of 10--20 LAEs and a binned average of 41 LBGs. 

We explain the anti-correlation of \dvlya\ and \wlya\ as a natural consequence of a change in the column density, covering fraction or velocity dispersion of gas near the systemic velocity. Increasing any of these quantities will require that a \lya\ photon acquire a larger frequency shift in order to escape the galaxy, thereby both shifting its ultimate velocity to redder wavelengths and increasing the number of scatterings it undergoes before escape. An increased number of scatterings increases the probability of absorption by dust, and may also result in scattering outside the aperture of the spectroscopic slit (e.g.\ \citealt{sbs+11}). Thus a shift to higher velocities is naturally accompanied by a decrease in the observed strength of the line (see also \citealt{jess13}).

An increase in the amount of gas at the systemic velocity may be related to the growth of galactic disks. \citet{lss+12b} find that galaxies with strong \lya\ emission tend to be compact or have multiple components, and suggest that large, rotationally-supported gaseous disks may develop later, after a sufficiently massive stellar component is in place to stabilize them. This larger ISM component would then both decrease the strength of \lya\ emission and shift it to the red, as described above. A related effect may be the harder ionizing spectrum of young, low metallicity stellar populations; \citet{lss+12b} also find that the outflows of smaller galaxies are more highly ionized, as traced by the relative strengths of the high and low ionization interstellar absorption lines. An increase in the ionization state of gas in and around the galaxy may also aid the escape of \lya\ photons, as previously suggested by \citet{eps+10} and \citet{hbo+11}. 

We comment on two further implications of the relationship between \wlya\ and \dvlya. First, our result in Section \ref{sec:dv} that LAEs with ${\cal R} < 25.5$ have velocity offsets that are statistically indistinguishable from those of continuum-selected galaxies in the same magnitude range is at odds with the results of \citet{son+14}, who find a highly significant difference in \dvlya\ between their sample of 22 bright LAEs and 41 similarly bright continuum-selected galaxies from \citet{ses+10}.  The explanation for this difference is likely to be found in the \lya\ equivalent width distributions of the two samples. The mean photometric \wlya\ for our LAEs with ${\cal R} < 25.5$ is 33 \AA, with a range of 21 to 47 \AA; in contrast, nearly all of the LAEs considered by \citet{son+14} have \wlya~$>50$ \AA, and typical values are $\sim80$--100 \AA. Given the strong anti-correlation between \wlya\ and \dvlya, it is not surprising that the LAEs in our sample have larger velocity offsets than those of \citet{son+14}. We conclude that \dvlya\ is closely related to both galaxy luminosity and \lya\ equivalent width, and therefore comparisons between continuum and \lya-selected galaxies must take the distributions of both of these quantities into account. 

Second, the anti-correlation between \wlya\ and \dvlya\ has potential implications for the results of \citet{vdb+12}, who combine radiative transfer with hydrodynamical models of galaxy evolution to find that the small-scale structure of the ISM has a dominant effect on the emergent \lya\ profile. For a more realistic simulation in which most young stars are embedded in dense clouds, the \lya\ profile is strongly dependent on the inclination at which the galaxy is observed, with emission much stronger in face-on galaxies. \citet{son+14b} find some observational support for this scenario in the observation that LAEs with higher equivalent widths tend to have smaller ellipticity.  Locally and at redshifts up to $z\sim1$, galactic outflows are observed to be collimated perpendicular to galactic disks (e.g.\ \citealt{ksm+12,msc+12, rpk+13}), resulting in an increase in the velocity of the outflow when the disk is observed face-on. Under the assumption that the velocity offset of \lya\ emission reflects the velocity of the outflow, one would then expect an {\it increase} in \dvlya\ with increasing equivalent width, the opposite of what is observed. There are at least two resolutions to this apparent contradiction:\ first, as discussed above, the velocity of \lya\ emission may not be driven primarily by the velocity of outflowing gas, and second, outflows in galaxies at $z\gtrsim2$ may not yet be strongly collimated \citep{lss+12b}. Simulations including both realistic feedback prescriptions and \lya\ radiative transfer for a variety of galaxies may clarify the relationship between galactic outflows and \lya\ emission, as will additional observations of \lya\ emission in galaxies with a wide range of outflow morphologies.

\subsection{Implications for the escape of Lyman continuum photons}

The same factors that allow \lya\ photons to escape near the systemic velocity of a galaxy, a low covering fraction and/or a low column density of neutral hydrogen, may also aid the escape of Lyman continuum (LyC) photons.  Evidence for this is seen in the higher LyC escape fractions of LAEs compared to LBGs \citep{nsss11,nsk+13,msn+13}. A spectroscopic study of LyC emission in continuum-selected galaxies at $z\sim3$ also finds that  \lya\ emission is stronger and has a higher blue fraction in galaxies with LyC emission relative to those without (C.\ Steidel et al.\ in prep). The simultaneous increase of both \lya\ and LyC emission cannot continue to the highest LyC escape fractions, however; if the LyC escape fraction is high, as in a density-bounded nebula, \lya\ emission will be weaker as few ionizing photons are converted to \lya\ photons. This effect may already be visible in LAE-based Lyman continuum studies, in which LyC-detected galaxies are seen to have lower \lya\ equivalent widths than LyC non-detections \citep{nsss11,msn+13}. 

While the relationship between \lya\ and LyC emission is not yet fully understood, a promising avenue in the search for galaxies with significant LyC emission may be the targeting of galaxies with strong and relatively symmetric \lya\ profiles, including substantial emission emerging on the blue side of zero velocity.  Our low spectral resolution makes this suggestion difficult to test precisely, but we can compare the \lya\ and LyC properties for the $z\sim3$ SSA22 LAEs in our sample, since they are drawn from the narrow-band imaging sample of \citet{nsss11}, who used a custom NB3640 filter to target LyC emission. If \lya\ and LyC emission are related as suggested above, we might expect the galaxies with the largest values of \Alya\ to be the most likely to be detected with LyC imaging. As in Section \ref{sec:asym} above, we consider only objects with uncertainties in \Alya\ less than 0.3; this includes 16 of the 19 SSA22 LAEs. This test yields mixed results:\ three of the LAEs in our sample are detected in the NB3640 filter (SSA22-003, SSA22-021, and SSA22-046), and one of these, SSA22-003, has the highest value of \Alya\ in the sample, with \Alya~$=0.95\pm0.07$. This is the only object in the sample with a \lya\ profile consistent with perfect symmetry, in which \Alya~$=1$.  However, the other two objects with NB3640 detections are unremarkable, with nothing in their \lya\ profiles to distinguish them from the rest of the LAE sample. 

We conclude that, while higher spectral resolution is clearly required in order to obtain a complete understanding of the relationship between LyC and \lya\ emission, targeting galaxies for LyC studies based on their \lya\ profiles may be promising, and will improve our knowledge of the link between LyC and \lya\ emission regardless of whether or not LyC emission is actually detected. 

We also note that recent radiative transfer modeling supports this suggestion. \citet{vosh14} have modeled the \lya\ profiles of galaxies which allow the escape of LyC photons, either because one or more of their \HII\ regions is density-bounded or because they have a non-unity covering fraction of neutral hydrogen. In both cases, they find distinctive features in the \lya\ profile:\ in the first case, the line is redshifted with a small velocity offset relative to systemic, \dvlya~$<150$ \kms, while in the second case there is a primary peak at the systemic redshift, which also results in significant \lya\ flux emerging blueward of zero velocity.  They therefore suggest that, given high resolution spectra, the \lya\ profile may be used to identify galaxies likely to have escaping LyC emission.  

\subsection{Future prospects}

There are many future observations that will clarify the results presented here. Perhaps most importantly, absorption-line studies of faint galaxies, either from very deep spectra or from larger telescopes, will provide constraints on the outflow velocities of these objects, giving insight into the relationship between \dvlya\ and outflow velocity in faint objects. Observations of larger samples of LAEs in a diversity of environments are also required in order to put the current results in context; all of the LAEs in our sample are found in overdense regions, and earlier work has suggested that the \lya\ properties of galaxies may depend on environment, with more spatially extended \lya\ profiles found in denser regions \citep{myh+12}.

Higher resolution spectroscopy of \lya\ emission will enable much more detailed modeling of the line profile, and the results of such modeling can be compared to additional and more detailed observations of the distribution, kinematics and physical conditions of the gas. While high resolution spectroscopy is likely to remain difficult for all but the most luminous (or gravitationally lensed; see \citealt{qpss09} for an example) \lya-emitting galaxies until the advent of larger telescopes, improved constraints on the physical conditions of the gas in high redshift \lya-emitting galaxies will come much sooner, from large samples of the rest-frame optical nebular emission lines measured from new near-IR multi-object spectrographs such as MOSFIRE.  

All of these studies will improve our understanding of the relationship between \lya\ emission and the physical conditions in and around galaxies, and we expect that this understanding will become increasingly valuable at higher redshifts, when \lya\ emission is often the only spectral information available. 

\acknowledgements

We would like to thank the referee for a thoughtful and constructive report. DKE is supported by the US National Science Foundation through the Faculty Early Career Development (CAREER) Program, grant AST-1255591. Additional support comes from the NSF through grants AST-0908805 (CCS, GCR, MB) and AST-1313472 (CCS, RFT, ALS), and an NSF Graduate Student Research Fellowship (ALS).  MB acknowledges support of Serbian MESTD through grant ON176021. MOSFIRE was made possible by grants to WMKO from the NSF ``Telescope System Instrumentation Program" (TSIP) and a generous donation from Gordon and Betty Moore. We thank our colleagues on the MOSFIRE instrument team, particularly Marcia Brown, Khan Bui, John Cromer, Jason Fucik, Hector Rodriguez, Bob Weber, and Jeff Zolkower at Caltech, Ted Aliado, George Brims, John Canfield, Chris Johnson, Ken Magnone, and Jason Weiss at UCLA, Harland Epps at UCO/Lick Observatory, and Sean Adkins at WMKO. Special thanks to all of the WMKO staff who helped make MOSFIRE commissioning successful, especially Marc Kassis, Allan Honey, Greg Wirth, Shui Kwok, Liz Chock, and Jim Lyke. Finally, we wish to extend thanks to those of Hawaiian ancestry on whose sacred mountain we are privileged to be guests.

\bibliographystyle{apj}

\end{document}